\documentclass[prd,a4paper,nofootinbib,
notitlepage,
]{revtex4-1}
\pdfoutput=1
\usepackage{graphicx}
\usepackage{bbold}
\usepackage{mathtools}
\usepackage{amsmath}
\usepackage{amsfonts}
\usepackage{amssymb}
\usepackage{amsbsy}
\usepackage{physics} 
\usepackage{bbding}
\usepackage{slashed}
\usepackage{color}
\usepackage{tabularx}
\usepackage{hyperref} 
\usepackage{orcidlink}
\usepackage{ulem}
\usepackage{bm}
\usepackage{mathrsfs}

\definecolor{oucrimsonred}{rgb}{0.6, 0.0, 0.0}
\definecolor{DarkGray}{gray}{0.4}
\definecolor{Gray}{gray}{0.6}
\definecolor{forestgreen}{rgb}{0.13,0.35,0.13}
\definecolor{ocre}{HTML}{F16723}
\definecolor{blueline}{rgb}{0,0.27,0.55}
\definecolor{persianblue}{rgb}{0.11, 0.22, 0.73} 
\colorlet{grayline}{gray!70}

\newcommand{\on}[1]{\operatorname{#1}}

\newcommand{\ii}{\mathbf{1}}

\def\eq#1{{Eq.~(\ref{#1})}}

\def\abs#1{\left| #1\right|}

\hypersetup{colorlinks, citecolor=forestgreen, linkcolor=forestgreen, urlcolor=forestgreen}
\newcommand{\be}{\begin{equation}}
\newcommand{\ee}{\end{equation}}
\newcommand{\bea}{\begin{eqnarray}}
\newcommand{\eea}{\end{eqnarray}}
\newcommand{\nn}{\nonumber}
\newcommand{\com}[1]{}
\newcommand{\gsim}{\lower.7ex\hbox{$\;\stackrel{\textstyle>}{\sim}\;$}}
\newcommand{\lsim}{\lower.7ex\hbox{$\;\stackrel{\textstyle<}{\sim}\;$}}

\begin{document}

\hypersetup{citecolor = forestgreen,
linktoc = section,
linkcolor = forestgreen,
urlcolor = forestgreen
}

\title[]{ \Large \color{oucrimsonred} \textbf{
A study of entanglement among particle momenta\\[-2pt] \large{I, $\tau$-lepton pairs at Belle II}
} }
\author{\bf M. Fabbrichesi$^{a\, \orcidlink{0000-0003-1937-3854}}$}
\author{\bf   R. Floreanini$^{a}\, \orcidlink{0000-0002-0424-2707}$}
\author{\bf L. Marzola$^{{b,c\, \orcidlink{0000-0003-2045-1100}}}$}
\affiliation{$^{a}$INFN, Sezione di Trieste, Via Valerio 2, I-34127 Trieste, Italy}
\affiliation{$^{b}$Laboratory of High-Energy and Computational Physics, NICPB, R\"avala pst 10, 10143 Tallinn, Estonia}
\affiliation{$^{c}$Institute of Computer Science, University of Tartu, Narva mnt 18, 51009 Tartu, Estonia}

\begin{abstract}
\noindent We study the entanglement of the momenta of particle pairs produced at colliders, focusing on $\tau$-lepton pairs at Belle~II. We reduce the momenta to qubits in two ways: by projecting onto hemisphere states, which keep only a part of the produced state, and with pseudo-spin operators, built from the hemisphere projectors and the inversion of the momenta, which keep the whole state. We also construct the joint state of momenta and spins, which depends on how the spin is defined at different momenta. When the spin is quantized along a fixed axis, parity ensures that the momenta are in a maximally entangled pure state, which is not entangled with the spins.  In the helicity basis the hemisphere states show a pair entangled in momentum and in spin at the same time, and the mass-suppressed amplitudes entangle the momenta with the spins, which accounts for the resulting mixed momentum state. These are all properties of particle states that are useful to know, even though momentum entanglement cannot be directly tested at colliders because all the measurable observables commute with the momenta, and no complementary measurement is available.
\end{abstract}

\maketitle

\section{Introduction}

The most distinctive and unexpected phenomena in quantum mechanics can be traced back to the superposition principle. When multiple particles are considered, the superposition of their possible states leads to entanglement~\cite{Horodecki:2009zz,Benatti:2010,Nielsen:2012yss,bruss2019quantum}---a property for which observables such as momentum, position, and spin cannot be uniquely assigned to each particle independently of the others. Note that analyzing the entanglement between two particles always involves observables that commute and act on each particle separately, thereby making the two particles separate subsystems.

The momentum entanglement discussed here is conceptually different from the helicity (spin) entanglement that is more commonly encountered in the literature related to quantum mechanics effects at colliders (see, for instance, the reviews~\cite{Barr:2024djo,Afik:2025ejh}). The presence of entanglement in the helicity state of particle pairs has been experimentally demonstrated by using the decays of $B$ mesons~\cite{Fabbrichesi:2023idl}, top-quark pair production~\cite{ATLAS:2023fsd,CMS:2024pts} and in the decays of the charmonium system~\cite{Fabbrichesi:2024rec}. In all these analyses, entanglement in the spin degrees of freedom is computed or reconstructed as a function of the scattering angle of each event. In this paper, we first discard information about the spin state of the analyzed particle pairs and evaluate the presence of entanglement between their momentum states. After that, we retain the spins of the produced particles and also study the joint state of their momenta and spins.

What distinguishes the momenta from the spins (apart from their commutation relations)  is that they are continuous variables, which must be discretized before the tools developed for qubits can be applied, and that at colliders they are measured only through the direction of each particle.
Particles originating from a decay or a collision exist in a superposition of their momentum states. This can be most readily understood by considering a particle produced in a decay. For a generic spherically symmetric interaction, the particle is in an $S$-wave state. Consequently, its state is a superposition of the momentum eigenstates pointing in all possible directions. When two or more particles are involved in a process, superposition can lead to entanglement.

The related question about how spherically symmetric $S$-waves transform into the linear tracks observed in a detector was answered long ago by Mott's work~\cite{Mott1929} (for a review, see \cite{Figari2014}). The entire process is governed uniquely by the Schr\"odinger equation in which the electrons of the atoms making up the detector appear in the perturbative expansion at the first and higher orders. This description shows that---at least in particle physics---quantum states can be thought of as always evolving deterministically, eliminating the need for measurement to create the mixed states that are ultimately measured~\cite{RevModPhys.75.715}.

\subsection{Position--momentum correlations and the Reid criterion}

The most direct way of probing quantum correlations in phase space would be to follow the original formulation of the Einstein--Podolsky--Rosen (EPR) paradox~\cite{Einstein:1935rr}, in terms of the positions and momenta of the two particles. Since the variables are continuous, entanglement can be tested by means of inequalities involving the dispersions of their values (see~\cite{Reid:2009zz,Patil:2025jvf,Giese:2025rbx} for reviews). Given the difference $x_-=x_2-x_1$ of the coordinates of the two particles along a chosen direction, and the sum $p_+=p_1+p_2$ of the corresponding momenta, one considers the dispersions
\be
\Delta x_- = \sqrt{\langle x_-^2 \rangle - \langle x_- \rangle^2}\, , \qquad
\Delta p_+ = \sqrt{\langle p_+^2 \rangle - \langle p_+ \rangle^2}\, .
\ee
Although $[x_i,\,p_j]=i\,\delta_{ij}$, the variables $x_-$ and $p_+$ commute and can be simultaneously sharp. In natural units, $c=\hbar=1$, every separable state satisfies~\cite{Mancini:2001qbv}
\be
\Delta x_-\, \Delta p_+ \geq 1\, , \label{eq:delta}
\ee
so that a violation of \eq{eq:delta} along at least one direction witnesses entanglement. Moreover, a product $\Delta x_-\, \Delta p_+ < 1/2$ satisfies the EPR-paradox criterion of Reid~\cite{PhysRevA.40.913,PhysRevLett.60.2731,PhysRevLett.68.3663}, signaling a form of nonlocality. These criteria have been tested in optical experiments~\cite{PhysRevLett.92.210403}.

Although appealing, because it involves variables with a well-understood classical meaning, this test cannot be carried out at colliders. The bounds in \eq{eq:delta} are set by the single-particle commutators and are therefore expressed in units of $\hbar$, whereas the dispersions of the positions and momenta reconstructed at colliders are macroscopic. Consider, for instance, $\tau$-lepton pairs produced at Belle~II, $e^+e^-\to\tau^+\tau^-$ at $\sqrt s=10.58$ GeV, with $\tau^\pm\to\pi^\pm\nu$, taking as coordinates the secondary vertices where the pions are created and as momenta those of the pions. The $\tau$ decay length is exponentially distributed, with mean $\beta\gamma c\tau_\tau\simeq 2.8\times 87\ \mu\text{m}\simeq 0.25$ mm, so that $\Delta x_-$ is a few tenths of a millimeter. The two neutrinos carry away a variable fraction of the $\tau$ momenta, so that $\Delta p_+$ is of the order of a GeV. A simple leading-order simulation of $e^+e^-\to\tau^+\tau^-\to\pi^+\pi^-\nu_\tau\bar\nu_\tau$, with exponentially distributed decay lengths and isotropic $\tau\to\pi\nu$ decays (neglecting spin correlations), gives, for the direction that minimizes the product, $\Delta x_-\simeq0.33$ mm and $\Delta p_+\simeq1.3$ GeV, that is
\be
\Delta x_-\, \Delta p_+ \simeq 0.4\ \text{mm}\cdot\text{GeV}\, .
\ee
Since $\hbar c = 1.97\times 10^{-13}$ GeV$\,$mm, this corresponds to $\Delta x_-\, \Delta p_+ \simeq 2\times 10^{12}\, \hbar$, and \eq{eq:delta} is satisfied by about twelve orders of magnitude. The dispersions are determined by the decay kinematics, and detector effects play no role.

Nor can $\Delta x_-$ be replaced by an intrinsic length, such as the Compton wavelength $\hbar/(mc)$ of the particles. The criterion requires the dispersion of $x_-$ in the actual two-particle state, measured on the same ensemble as $\Delta p_+$. With $\Delta x_-=\hbar/(mc)$ the product reduces to the kinematic ratio $\Delta p_+/m$, whose value depends on the arbitrary choice of mass: about 9 for the pion and about 0.7 for the $\tau$ lepton. The Compton wavelength sets the scale below which a single relativistic particle cannot be localized: it is a lower bound on the position spread of each particle, not an estimate of the spread of $x_2-x_1$, which varies from event to event over the decay-length distribution. Notice that the spread of the relative position cannot be discarded as classical: the exponential distribution of decay times is itself a property of the quantum state, and the criterion involves its full variance. Even the experimental resolutions alone, of order $10\ \mu$m for vertices and MeV for momenta, would give a product of order $10^{8}\,\hbar$. The same conclusion holds for any choice of positions and momenta reconstructed in a collider detector: the scales involved are many orders of magnitude above $\hbar$, and position--momentum EPR correlations are out of reach.

In the following, we therefore start by considering the momenta alone. We discretize the problem by using hemispherical projectors, thereby defining dichotomic variables that map the two-particle momentum states onto the states of a bipartite qubit system. We employ these variables to test entanglement and Bell nonlocality using the concurrence and the Horodecki criterion (Sec.~\ref{sec:momentum}). In Sec.~\ref{sec:spin_momentum} we also consider the spins and construct the joint state of the discretized momenta and the helicities, a state of one momentum qubit and two spin qubits. This requires a definition of the spin at different momenta: the spin operators must commute with the momenta for the decomposition into momentum and spin to exist, but this condition does not select a unique spin description. We then compare the helicity basis with the canonical basis, in which the spin is quantized along a fixed axis. In the helicity basis, and in the massless limit, the $\tau$ pair is entangled in momentum and in helicity at the same time, a hyperentangled state. The mass-suppressed amplitudes then entangle the momenta with the spins, and this entanglement is what makes the momentum state mixed once the spins are traced out. In the canonical basis, parity forces the momentum state to be a Bell state and leaves momentum and spin unentangled. In Sec.~\ref{sec:pseudospin} we repeat the analysis with a different discretization based on pseudo-spin operators, which keep the whole of the produced state and yield the measured hemisphere probabilities on the diagonal of the density matrix. In Sec.~\ref{sec:measurements} we use these operators to show that none of the entanglement involving the momenta can be witnessed with collider measurements, which necessarily involve states with definite momenta.

\section{Momentum entanglement}
\label{sec:momentum}

Momentum entanglement in particle collisions has been discussed in various works~\cite{Seki:2014cgq,Peschanski:2016hgk,Peschanski:2019yah,Faleiro:2016lsf,Kowalska:2024kbs,Low:2024hvn,Sou:2025tyf,Peschanski:2026edo}, mostly in the context of entanglement between the forward and scattered particles. In our case, we only consider entanglement among the particles produced in the collision.

In order to utilize quantum information tools available for $n$-qubit systems, we convert the particle momentum into dichotomic variables by considering equivalence classes defined by hemispherical projections. The momentum of a particle can then be found in either one hemisphere or the other, and is thus mapped onto a qubit state which indicates the hemisphere it crosses. A discretization of continuous variables in the computation of quantum observables has been previously proposed in Refs.~\cite{Bell_Aspect_2004,Cetto1985} and recently implemented for momenta in the decays of the Higgs boson into gauge bosons at the LHC~\cite{Aguilar-Saavedra:2025njw}.

As a working example, we focus on the $\tau$-lepton pairs produced in electron--positron annihilation because the process is clean, the momenta of these particles can be fully reconstructed for a large fraction of their subsequent decays, and more than a million events have already been collected at Belle~II~\cite{Belle-II:2018jsg,Belle-II:2024vuc}.

\subsection{Two-particle states and the $S$ matrix}

We consider the following basis for two-particle states belonging to either initial- or final-state Fock spaces
\begin{equation}
	\ket{\alpha,\,p_1,\,s_1;\,\beta,\,p_2,\,s_2} = \ket{\alpha,\,p_1,\,s_1}\otimes \ket{\beta,\,p_2,\,s_2},
\end{equation}
where $\alpha$ and $\beta$ indicate the type of particle, and $p_i$ and $s_i$ are the momentum and helicity of the $i$th particle, respectively. The states are normalized as
\begin{align}
	\braket{\alpha,\,p_1,\,s_1;\,\beta,\,p_2,\,s_2}{\alpha',\,p'_1,\,s'_1;\beta',\,p'_2,\,s'_2} &= 4 E_{p_1}\,E_{p_2}\,\qty(2\pi)^3 \delta^3\qty(\vec{p}_1-\vec{p}_1{}')\,\qty(2\pi)^3 \delta^3\qty(\vec{p}_2-\vec{p}_2{}') \\ \nonumber &\qquad \times\delta_{s_1 s'_1} \delta_{s_2 s'_2}\,\delta_{\alpha \alpha'} \delta_{\beta\beta'},
\end{align}
and satisfy the completeness relation
\begin{equation}
	\sum_{\alpha,\beta,s_1, s_2} \int \prod_{i=1}^2\left( \frac{1}{2E_{p_i}} \frac{\dd^3p_i}{(2\pi)^3} \right)	\dyad{\alpha,\,p_1,\,s_1;\,\beta,\,p_2,\,s_2} = 1.
\end{equation}

Initial and final states are connected by the $\on{S}$ matrix
\begin{equation}
	\on S = \ii + i \on T,
\end{equation}
and, by specifying the initial state $\ket{\gamma,\,q_1,\,r_1;\,\delta,\,\,q_2,\,r_2}$, we obtain
\begin{align}
	\label{eq:outstate1}
	&
	\on{S} \ket{\gamma,\,q_1,\,r_1;\,\delta,\,q_2,\,r_2} \propto
	\ket{\gamma,\,q_1,\,r_1;\,\delta,\,\,q_2,\,r_2}
	\\ \nonumber
	&\qquad +
	\sum_{\alpha,\beta,s_1, s_2} \int \prod_{i=1}^2\left( \frac{1}{2E_{p_i}} \frac{\dd^3p_i}{(2\pi)^3} \right)	\mel{\alpha,\,p_1,\,s_1;\,\beta,\,p_2,\,s_2}{i\on{T}}{\gamma,\,q_1,\,r_1;\,\delta,\,\,q_2,\,r_2}\,\ket{\alpha,\,p_1,\,s_1;\,\beta,\,p_2,\,s_2}
	\\ \nonumber
	&\qquad= \ket{\gamma,\,q_1,\,r_1;\,\delta,\,\,q_2,\,r_2}+  i\sum_{\alpha,\beta,s_1, s_2} \int \dd\Pi_2(\vec p_1, \vec p_2)\, \mathcal{M}^{\alpha\beta\gamma\delta}_{s_1s_2r_1r_2}(p_1,\,p_2,\,q_1,\,q_2) \,\ket{\alpha,\,p_1,\,s_1;\,\beta,\,p_2,\,s_2},
\end{align}
where the proportionality factor is set by the normalization condition and in which we used the definition of the amplitude
\begin{equation}
	\mel{\alpha,\,p_1,\,s_1;\,\beta,\,p_2,\,s_2}{i\on{T}}{\gamma,\,q_1,\,r_1;\,\delta,\,\,q_2,\,r_2}
	= \qty(2\pi)^4 \delta(q_1+q_2-p_1-p_2)\,i\mathcal{M}^{\alpha\beta\gamma\delta}_{s_1s_2r_1r_2}(p_1,\,p_2,\,q_1,\,q_2).
\end{equation}
The factor enforcing the four-momentum conservation has been incorporated into the infinitesimal volume to give the Lorentz-invariant phase-space measure
\begin{equation}
\dd\Pi_2 = 	\qty(2\pi)^4 \delta(q_1+q_2-p_1-p_2) \prod_{i=1}^2\left( \frac{1}{2E_{p_i}} \frac{\dd^3p_i}{(2\pi)^3} \right).
\end{equation}

Restricting ourselves to the analysis of the $\tau$-pair production in electron--positron annihilation, we set $\alpha = \tau^+$, $\beta=\tau^-$, $\gamma =e^+$ and $\delta=e^-$ and simplify the formalism by suppressing these flavor indices henceforth. In order to select the final state of interest we act on \eq{eq:outstate1} with the projector on the $\tau^+\tau^-$ subspace, $\Pi_{\bar\tau\tau}$, obtaining
\begin{equation}
	\label{eq:outstate2}
	\ket{\Psi_{r_1r_2}^{\bar\tau \tau}(q_1, q_2)} =
	\Pi_{\bar\tau\tau}\on{S} \ket{q_1,\,r_1;\,q_2,\,r_2}
	\propto
	\sum_{s_1, s_2} \int \dd\Pi_2(\vec p_1, \vec p_2, \vec q_1, \vec q_2)\, \mathcal{M}_{s_1s_2r_1r_2}(p_1,\,p_2,\,q_1,\,q_2) \,\ket{p_1,\,s_1;\,p_2,\,s_2}.
\end{equation}

\subsection{Momentum density matrices}

Each initial state with polarized electron and positron prepares a pure $\tau$-pair state. Initial states with the same helicity yield momentum states proportional to the electron mass. In the following, we neglect them and consider only states obtainable with massless initial-state particles, as appropriate in energy regimes where the electron mass can be neglected. For the two configurations with opposite helicities, the states are
\begin{align}
\ket{\Psi_{-+}^{\bar\tau \tau}(q_1, q_2)}
	\propto&
	-ie^2\int \dd\Pi_2 \, e^{-i\phi} \qty[\qty(\cos\theta+1)\ket{p_1,\,-;\,p_2,\,+} + \qty(\cos\theta-1)\ket{p_1,\,+;\,p_2,\,-}]
	\\ \nonumber
	& +\frac{2 i e^2 m_\tau}{\sqrt{s}}\int \dd\Pi_2 \, \sin\theta \qty(\ket{p_1,\,-;\,p_2,\,-}-e^{-2 i \phi } \ket{p_1,\,+;\,p_2,\,+}),
	\\
\ket{\Psi_{+-}^{\bar\tau \tau}(q_1, q_2)}
	\propto&
	-ie^2 \int \dd \Pi_2\, e^{i\phi} \qty[\qty(\cos\theta-1)\ket{p_1,\,-;\,p_2,\,+} + \qty(\cos\theta+1)\ket{p_1,\,+;\,p_2,\,-}]
	\\ \nonumber
	& +\frac{2 i e^2 m_\tau}{\sqrt{s}}\int \dd \Pi_2 \sin\theta \qty(e^{2 i \phi }\ket{p_1,\,-;\,p_2,\,-}- \ket{p_1,\,+;\,p_2,\,+}),
\end{align}
where the $\pm$ indices in the state on the left-hand side correspond to a choice of $r_1$ and $r_2$, the helicities of the initial positron and electron, respectively, $\sqrt s$ is the center-of-mass energy, and $p_1$ and $p_2$ are the momenta of the $\tau^+$ and $\tau^-$. In our conventions, the scattering angle $\theta$ is taken between the directions of the $\tau^+$ and incoming $e^+$, while $\phi$ is the azimuthal angle. The helicity states are defined by the rotation $R(\phi_p,\theta_p,-\phi_p)$ that takes the $z$ axis into the direction of the momentum: the two-component spinors are $\chi_+=(\cos\frac{\theta_p}{2},\,e^{i\phi_p}\sin\frac{\theta_p}{2})$ and $\chi_-=(-e^{-i\phi_p}\sin\frac{\theta_p}{2},\,\cos\frac{\theta_p}{2})$, with $(\theta_p,\phi_p)$ the polar angles of the momentum of each lepton, $(\theta,\phi)$ for the $\tau^+$ and $(\pi-\theta,\phi+\pi)$ for the $\tau^-$. 

In order to discard information about the $\tau$-pair helicity state, we compute the corresponding density matrices and perform the partial trace on the helicity states $\ket{s_1, s_2}$ with $s_i=\pm$, obtaining the following (unnormalized) density operators on momentum space:
\begin{align}
	\rho_{-+} &= \tr_{s_1,s_2}\qty(\dyad{\Psi_{-+}^{\bar\tau \tau}})
	\nonumber \\
	& \propto e^4 \int \dd \Pi_2 \dd \Pi_2'\,
	\qty[2e^{-i(\phi-\phi')}\qty(\cos\theta\cos\theta'+1)
	+\frac{4 m_\tau^2}{s}\sin\theta\sin\theta'\qty(1+e^{-2i(\phi-\phi')})]
	\dyad{p_1\,;\,p_2}{p'_1\,;\,p'_2}, \label{eq:rhomp_cont}
	\\
	\rho_{+-} &= \tr_{s_1,s_2}\qty(\dyad{\Psi_{+-}^{\bar\tau \tau}})
	\nonumber \\
	& \propto e^4 \int \dd \Pi_2 \dd \Pi_2'\, \qty[2e^{i(\phi-\phi')}\qty(\cos\theta\cos\theta'+1)+\frac{4 m_\tau^2}{s}\sin\theta\sin\theta'\qty(1+e^{2i(\phi-\phi')})]
	\dyad{p_1\,;\,p_2}{p'_1\,;\,p'_2}. \label{eq:rhopm_cont}
\end{align}

The helicity states $\ket{p,\,s}$ at different momenta are different spin states, so the partial trace over helicities implicitly identifies the spin states at different momenta. The resulting momentum state depends on this choice, and on the phases of the helicity states fixed above. We use the helicity basis throughout this section, and discuss the alternative of a spin quantized along a fixed axis, and its consequences, in Sec.~\ref{sec:spin_momentum}.

\subsection{Coarse graining}

The density matrices $\rho_{-+}$ and $\rho_{+-}$ are infinite dimensional, making the study of momentum states highly non-trivial. A possible way to simplify the task is provided by coarse graining, in which the momentum of each particle is assigned to one of two hemispheres. Given an arbitrary direction $\hat n$ in the center-of-mass frame, the hemisphere $h=\mathfrak{f}$ is defined by $\vec k\cdot\hat n\geq0$ and $h=\mathfrak{b}$ by $\vec k\cdot\hat n<0$. In the following we take $\hat n$ along the $y$ or the $z$ axis, the $x$ axis being equivalent to $y$.

Each hemisphere is associated with the projector
\begin{equation}
	\label{eq:h_proj1p}
	\Lambda_{h} =\int\limits_{\text{supp}(h)}  \frac{1}{2E_{k}} \frac{\dd^3k}{(2\pi)^3} \dyad{k},\qquad
	\Lambda_h \ket{p} = \Theta_h(p) \ket{p},\qquad
	\Theta_h(p) =
	\begin{cases}
	  1 & \vec p \in h, \\
	  0 & \vec p \notin h,
	\end{cases}
\end{equation}
an infinite-rank projector that signals whether a particle is found in $h$. The probability $\tr[\Lambda_h\rho]$ is what a detector measures. The coarse graining is defined instead through the hemisphere state $\ket h$, the uniform superposition of the momentum eigenstates in $h$ (at the fixed magnitude of the momentum set by the kinematics), with overlaps
\begin{equation}
	\label{eq:overlpa_hp1}
	\braket{h}{p} = c\,\Theta_h(p),
\end{equation}
where $c$ is a normalization constant that drops out of the normalized states. Within the momentum space of a particle, the two hemisphere states span a two-dimensional subspace, and we associate with them the basis states $\ket{\mathfrak f}_{\rm q}$ and $\ket{\mathfrak b}_{\rm q}$ of a qubit. The two-particle hemisphere states are $\ket{h_1;\,h_2}=\ket{h_1}\otimes\ket{h_2}$, and we define the coarse-grained momentum state through its matrix elements on them, evaluated between the qubit states associated with them, $\ket{h_1;\,h_2}_{\rm q}=\ket{h_1}_{\rm q}\otimes\ket{h_2}_{\rm q}$,
\begin{align}
	\label{eq:p_projections}
	{}_{\rm q}\mel{h_1;\,h_2}{\rho_{-+}}{h_3;\,h_4}_{\rm q}
	&\propto e^4 \int\limits_{h_1h_2} \dd\Pi_2 \int\limits_{h_3h_4}\dd \Pi_2'\,
	\Bigg[2e^{-i(\phi-\phi')}\qty(\cos\theta\cos\theta'+1)
	+\frac{4 m_\tau^2}{s}\sin\theta\sin\theta'\qty(1+e^{-2i(\phi-\phi')})\Bigg],
\end{align}
and similarly for $\rho_{+-}$. 

Equivalently, the coarse-grained state is the image of the momentum state under the map $\rho\mapsto(V\otimes V)\,\rho\,(V\otimes V)^\dagger$, with $V=\sum_h\ket{h}_{\rm q}\!\bra{h}$ mapping each hemisphere state onto a qubit state $\ket{h}_{\rm q}$. The operator $V$ acts on the whole momentum space of the particle: it maps the normalized hemisphere states isometrically onto the qubit basis and sends to zero every state orthogonal to both of them, so that $V^\dagger V=\sum_h\ket{h}\!\bra{h}$ projects onto the two-dimensional subspace they span. Since $V\otimes V$ is a product of local operations, entanglement of the coarse-grained state implies entanglement of the momentum state. The map, however, does not preserve the trace, because the part of the produced state that is orthogonal to the hemisphere states is discarded. Moreover, the diagonal elements $\mel{h}{\rho}{h}$ coincide with the hemisphere probabilities $\tr[\Lambda_h\rho]$ only when the amplitudes are constant over each hemisphere. At fixed spin label, $\ket{\mathfrak f}+\ket{\mathfrak b}$ is the $S$ wave and $\ket{\mathfrak f}-\ket{\mathfrak b}$ contains only odd waves, three quarters of it being the $P$ wave along $\hat n$: the map keeps the part of the state that lies in these two waves and discards the rest. We call weight, $w$, the fraction of the produced state that is kept, that is, the trace of the coarse-grained state for normalized $\ket{h;\,\bar h}$ and $\tr\rho=1$, where $\bar h$ is the hemisphere opposite to $h$. In other words, the weight is the expectation value of the projector $V^\dagger V\otimes V^\dagger V$ on the produced state, $w=\tr[(V^\dagger V\otimes V^\dagger V)\,\rho]$. The coarse-grained state, and its off-diagonal elements in particular, are predictions of the $S$-matrix formalism, not measured quantities. In Sec.~\ref{sec:pseudospin} we consider a discretization that preserves the trace, and in Sec.~\ref{sec:measurements} we discuss what can actually be measured.

A representation of the qubit basis onto which $V$ maps the hemisphere states is provided by
\begin{equation}
	\label{eq:discr_repr_k}
	\ket{\mathfrak{f}}_{\rm q} = \mqty(1\\0), \quad\ket{\mathfrak{b}}_{\rm q} = \mqty(0\\1),
\end{equation}
and their Kronecker products,
\begin{equation}
	\ket{\mathfrak{f};\,\mathfrak{f}}_{\rm q} = \mqty(1\\0\\0\\0),
	\quad
	\ket{\mathfrak{f};\,\mathfrak{b}}_{\rm q} = \mqty(0\\1\\0\\0),
	\quad
	\ket{\mathfrak{b};\,\mathfrak{f}}_{\rm q} = \mqty(0\\0\\1\\0),
	\quad
	\ket{\mathfrak{b};\,\mathfrak{b}}_{\rm q} = \mqty(0\\0\\0\\1).
\end{equation}
From now on we omit the subscript, and the states $\ket{h_1;\,h_2}$ that label the coarse-grained state are understood to be these qubit states.

As the delta functions contained in the phase-space measures select only states where the two particles are back to back, we then have
\begin{equation}
	\label{eq:rhopm_p}
	\rho_{\pm\mp}
	\propto
	\mqty(
		0 &0 &0 &0\\
		0 & \mathcal{E}^{\pm\mp}_{\mathfrak{fb};\mathfrak{fb}}& \mathcal{E}^{\pm\mp}_{\mathfrak{fb};\mathfrak{bf}}&0\\
		0 & \mathcal{E}^{\pm\mp}_{\mathfrak{bf};\mathfrak{fb}}& \mathcal{E}^{\pm\mp}_{\mathfrak{bf};\mathfrak{bf}}&0\\
		0 &0 &0 &0),
\end{equation}
where we introduced the abbreviation $\mathcal{E}^{\pm\mp}_{h_1h_2;h_3h_4} = \mel{h_1;\,h_2}{\rho_{\pm\mp}}{h_3;\,h_4}$. In the center-of-mass frame $E_{p_{1,2}}=E_{p'_{1,2}} = \sqrt s /2$, and momentum conservation sets $h_2$ and $h_4$ as the hemispheres opposite to $h_1$ and $h_3$. The phase-space integrals in \eq{eq:p_projections} then reduce to integrals over the direction of $\vec p_1$, and
\begin{equation}
	\mathcal{E}^{\pm\mp}_{h_1h_2;h_3h_4}
	\propto
	\int\limits_{h_1} \dd\Omega_{p_1} \int\limits_{h_3}\dd \Omega_{p_1'}\,
	\Bigg[2e^{\pm i(\phi-\phi')}\qty(\cos\theta\cos\theta'+1)
	+\frac{4 m_\tau^2}{s}\sin\theta\sin\theta'\qty(1+e^{ \pm 2i(\phi-\phi')})\Bigg],
\end{equation}
where all constant factors have been absorbed into the normalization of the trace.

For mixed initial states, the two density matrices in \eq{eq:rhopm_p} are to be combined proportionally to the statistical weights characterizing the ensemble we aim to describe. For unpolarized electron beams we have
\begin{equation}
	\label{eq:rho_unpol}
	\rho = \frac{\rho_{-+}+\rho_{+-}}{2} =
	\mqty(
		0 &0 &0 &0\\
		0 & \mathcal{E}_{\mathfrak{fb};\mathfrak{fb}}& \mathcal{E}_{\mathfrak{fb};\mathfrak{bf}}&0\\
		0 & \mathcal{E}_{\mathfrak{bf};\mathfrak{fb}}& \mathcal{E}_{\mathfrak{bf};\mathfrak{bf}}&0\\
		0 &0 &0 &0),
\end{equation}
where we defined $\mathcal{E}_{h_1\bar h_1;h_3 \bar h_3} = \qty(\mathcal{E}^{+-}_{h_1\bar h_1;h_3 \bar h_3}+\mathcal{E}^{-+}_{h_1\bar h_1;h_3 \bar h_3})/2$ and again $\bar h$ denotes the hemisphere opposite to $h$. Explicitly, abbreviating $\dd\Omega_{p_1}$ with $\dd \Omega$ and $\dd\Omega_{p_1'}$ with $\dd \Omega'$, we have
\begin{equation}
	\label{eq:rho_mat_el_p}
	\mathcal{E}_{h_1\bar h_1;h_3 \bar h_3}
	\propto
	\int\limits_{h_1} \dd\Omega \int\limits_{h_3}\dd \Omega'\,
	\Bigg[2\cos(\phi-\phi')\qty(\cos\theta\cos\theta'+1)
	+R\,\sin\theta\sin\theta'\qty[1+\cos(2(\phi-\phi'))]\Bigg],
\end{equation}
having defined the parameter $R=4 m_\tau^2/s$, such that $0\leq R<1$.

\subsection{Quantum information observables}

The quantum information theory (QIT) quantities of interest are the purity of the state, computed as $\tr(\rho^2)$, the concurrence
\be
\mathcal{C}[\rho] = \max (0,\, r_1-r_2-r_3-r_4)\, ,
\ee
where $r_{i}$ are the square roots, in decreasing order, of the eigenvalues of the matrix
\be
{\cal R}=\rho (\sigma_{y}\otimes \sigma_{y}) \rho^{*} (\sigma_{y}\otimes \sigma_{y})\, .
\ee
In order to quantify Bell nonlocality, we adopt the Horodecki criterion
\be
\mathfrak{m}_{12} = m_{1}+m_{2} >  1\, ,
\ee
in which $m_{1}$ and $m_{2}$ are the two largest eigenvalues of the matrix
\be
M=C\,C^{T}\, ,
\ee
and $C$ is the correlation matrix of the Fano decomposition
\be
\rho=\frac14\Big[\mathbb{1}\otimes\mathbb{1}+\sum_i B_i^+\,\sigma_i\otimes\mathbb{1}+\sum_j B_j^-\,\mathbb{1}\otimes\sigma_j+\sum_{ij}C_{ij}\,\sigma_i\otimes\sigma_j\Big].
\ee
More about the use of these QIT quantities in collider settings can be found in the review~\cite{Barr:2024djo}.

\subsection{Coarse graining along the $z$ direction}

We take here $\text{supp}(h) = k_z\geq 0$ if $h = \mathfrak{f}$, and $k_z< 0$ if $h=\mathfrak{b}$. The integration domains in \eq{eq:rho_mat_el_p} are then $\theta\in[0, \pi/2]$, $\phi\in[0,2\pi]$ for $\mathfrak{f}$ and $\theta\in[\pi/2, \pi]$, $\phi\in[0,2\pi]$ for $\mathfrak{b}$.

\paragraph{Diagonal events.}
As a reference, we first remove all coherence between different momenta, keeping only the diagonal of the continuous momentum density matrix. This sets $\vec p_1 = \vec p_1\,'$, implying $\theta=\theta'$ and $\phi=\phi'$, and describes the classical ensemble with the same angular distribution as the quantum state; its diagonal elements are the measured hemisphere probabilities $\tr[\Lambda_h\rho]$. It yields the matrix elements
\begin{equation}
	\mathcal{E}_{h_1\bar h_1;h_1 \bar h_1}
	\propto
	\int\limits_{h_1} \dd{\cos\theta}
	\qty(\cos^2\theta
	+R\sin^2\theta+1).
\end{equation}
The corresponding density matrix is
\begin{equation}
	\label{eq:rhoadiag_ev}
	\rho	=
		\mqty(0 & 0 & 0 & 0\\
				 0 & \dfrac 12 & 0 & 0 \\
				 0 &0 & \dfrac12& 0\\
				 0 & 0 & 0 & 0),
\end{equation}
giving the following Fano coefficient decomposition:
\begin{equation}
	\vec B^+ = \vec B^- = \vec 0, \quad C = \mqty(0 & 0 & 0 \\ 0 & 0 & 0 \\ 0 & 0 & -1 ).
\end{equation}
The state is an incoherent mixture with equal weights, implying that each particle, $\tau^+$ or $\tau^-$, has equal probability to be found in either hemisphere. This is expected as the differential cross section for the process is symmetric in the scattering angle. The QIT quantities for the density matrix in \eq{eq:rhoadiag_ev} are
\begin{equation}
	\tr(\rho^2) = \frac12,
	\quad
	\mathcal{C}[\rho]  = 0,
	\quad
	\mathfrak{m}_{12} = 1,
\end{equation}
as characteristic of a classically correlated and local state.

\paragraph{Coplanar events.}
As an intermediate case, we remove the coherence in the azimuthal angle only, setting $\phi=\phi'$ in the integrand while keeping $\theta$ and $\theta'$ independent. This describes the state obtained when the azimuth, that is, the orientation of the scattering plane, is treated as classically known, as in the usual computation in which the scattering plane is taken to coincide with the $xz$ coordinate plane ($\phi=\phi'=0$). In this case the density matrix element in \eq{eq:rho_mat_el_p} simplifies to
\begin{equation}
	\mathcal{E}_{h_1\bar h_1;h_3 \bar h_3}
	\propto
	\int\limits_{h_1} \dd\cos\theta \int\limits_{h_3}\dd \cos\theta'\,
	\qty(\cos\theta\cos\theta'
	+R\sin\theta\sin\theta'+1),
\end{equation}
after performing the trivial azimuthal integrations. The remaining integrations give
\begin{align}
	\mathcal{E}_{\mathfrak{fb};\mathfrak{fb}}
	=
	\mathcal{E}_{\mathfrak{bf};\mathfrak{bf}}
	&\propto \frac{1}{16}\qty(20 + \pi^2 R),
	\\
	\mathcal{E}_{\mathfrak{fb};\mathfrak{bf}}
	=
	\mathcal{E}_{\mathfrak{bf};\mathfrak{fb}}
	&\propto \frac{1}{16}\qty(12 + \pi^2R),
\end{align}
and the normalized density matrix then is
\begin{equation}
	\label{eq:rhoacoplanar}
	\rho	=
		\mqty(0 & 0 & 0 & 0\\
				 0 & \dfrac 12 & \dfrac12 - \dfrac{4}{20 + \pi^2 R} & 0 \\
				 0 & \dfrac12 - \dfrac{4}{20 + \pi^2 R} & \dfrac12& 0\\
				 0 & 0 & 0 & 0)\xrightarrow{R\to0 \,(s\gg m_\tau^2)}
				 \mqty(0 & 0 & 0 & 0\\
				 0 & \frac 12 & \frac{3}{10} & 0 \\
				 0 & \frac{3}{10} & \frac12& 0\\
				 0 & 0 & 0 & 0).
\end{equation}
As explained before, that the diagonal elements equal 1/2 shows that each particle, $\tau^+$ or $\tau^-$, has equal probability to be found in either hemisphere. On the contrary, if we were to restrict the calculation to a specific $\tau$-pair spin state, we would observe that the diagonal terms do indeed differ. For instance, the $\ket{\mp \pm}$ states would give integrands equal to $(\cos\theta\pm1)(\cos\theta'\pm1)$ and diagonal elements of 1/10 and 9/10, to be compared with the fractions 1/8 and 7/8 of the events found in the two hemispheres. This again mirrors the results for the corresponding helicity amplitudes, which yield mirror-image distributions for the cross section plotted as a function of the scattering angle. The resulting forward--backward asymmetry therefore only vanishes once the two spin contributions are summed.

\begin{figure}[h!]
	\centering
	\includegraphics[width=0.55\linewidth]{"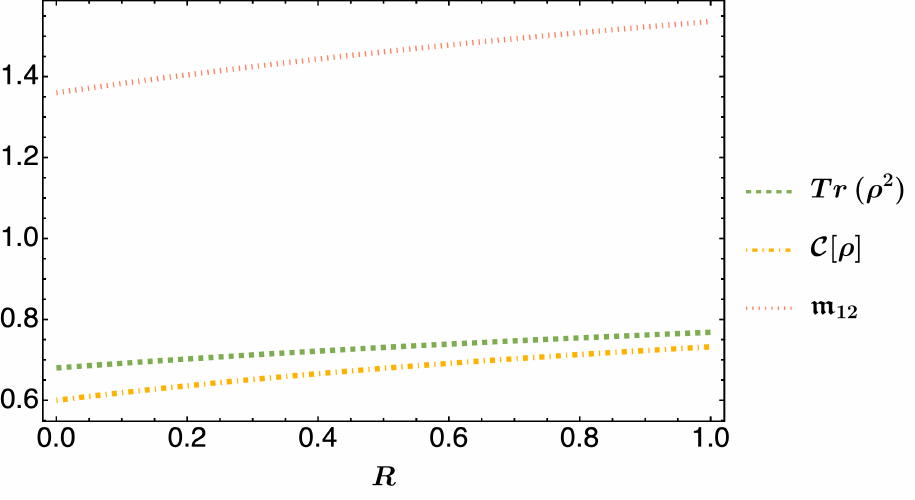"}
	\caption{Dependence of the indicated QIT quantities on the ratio $R=4m_\tau^2/s$ for the coplanar events with hemispheres defined along the $z$ direction.}
	\label{fig:coplanarQIT}
\end{figure}

The density matrix in \eq{eq:rhoacoplanar} is characterized by the following Fano coefficients
\begin{equation}
	\vec B^+ = \vec B^- = \vec 0, \quad C = \mqty(1 - \frac{8}{20+\pi^2 R} & 0 & 0 \\ 0 & 1 - \frac{8}{20+\pi^2 R} & 0 \\ 0 & 0 & -1 ),
\end{equation}
and, investigating the usual QIT quantities, we find
\begin{align}
	\tr(\rho^2) &= \frac12 + 2\qty(\frac 12 -\frac{4}{20 + \pi^2 R })^2,
	\\
	\mathcal{C}[\rho] &= 1 - \frac{8}{20 + \pi^2 R},
	\\
	\mathfrak{m}_{12} & = 1 + \qty(\frac{12+\pi^2 R}{20 + \pi^2 R})^2,
\end{align}
which are plotted in Fig.~\ref{fig:coplanarQIT} as functions of $R$. The state is an entangled and nonlocal mixture.

\paragraph{All events.}
Keeping the full coherence, the azimuthal integrations in \eq{eq:rho_mat_el_p} remove the first term, which comes from the helicity-conserving amplitudes, and we obtain
\begin{equation}
	\mathcal{E}_{h_1\bar h_1;h_3 \bar h_3}
	\propto R
	\int\limits_{h_1} \dd\cos\theta \sin\theta  \int\limits_{h_3}\dd\cos\theta'\,
	\sin\theta'\propto R\,\frac{\pi^2}{16}\quad\forall h_1, h_3.
\end{equation}
The corresponding density matrix then is
\begin{equation}
	\label{eq:rhoz}
	\rho	=
		\frac 12 \mqty(0 & 0 & 0 & 0\\
				 0 & 1 & 1 & 0 \\
				 0 &1 & 1& 0\\
				 0 & 0 & 0 & 0) \equiv \dyad{\Psi^+},
\end{equation}
where
\begin{equation}
	\ket{\Psi^+} = \frac{1}{\sqrt 2} \qty(\ket{\mathfrak{f};\mathfrak{b}}+\ket{\mathfrak{b};\mathfrak{f}}).
\end{equation}
The state is a Bell state, hence we have
\begin{equation}
	\tr(\rho^2)=1, \quad \mathcal{C}[\rho]=1, \quad \mathfrak{m}_{12} = 2\,,
\end{equation}
that is, the state is pure, maximally entangled and Bell nonlocal. The whole state comes from the helicity-flip amplitudes, suppressed by $2m_\tau/\sqrt s$: the weight of the coarse-grained state is
\begin{equation}
	\label{eq:weight_z}
	w = \frac{3\pi^2 R}{64\,(2+R)},
\end{equation}
equal to $0.025$ at the Belle~II energy, and vanishes in the massless limit, in which the hemisphere states along the beam do not overlap with the produced state.

\subsection{Coarse graining along the $y$ direction}

\begin{figure}[h!]
	\centering
	\includegraphics[width=0.6\linewidth]{"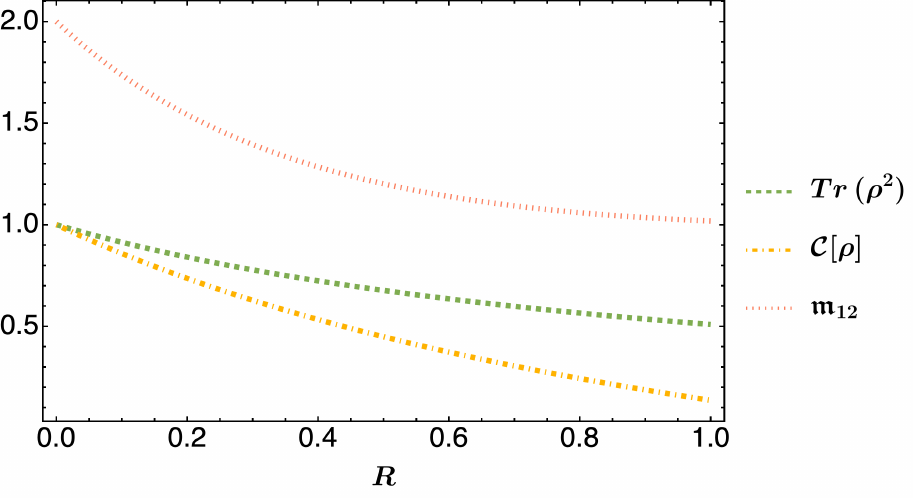"}
	\caption{Dependence of the indicated QIT quantities on the ratio $R=4m_\tau^2/s$ for the case of the hemispheres defined along the $y$ direction.}
	\label{fig:plotphiEnt}
\end{figure}

We take here $\text{supp}(h) = k_y\geq 0$ if $h = \mathfrak{f}$, and $k_y< 0$ if $h=\mathfrak{b}$. The integration domains in \eq{eq:rho_mat_el_p} are then $\theta\in[0, \pi]$, $\phi\in[0,\pi]$ for $\mathfrak{f}$ and $\theta\in[0, \pi]$, $\phi\in[\pi,2\pi]$ for $\mathfrak{b}$. Performing the integrations on the indicated domains we obtain
\begin{align}
	\mathcal{E}_{\mathfrak{fb};\mathfrak{fb}}
	=
	\mathcal{E}_{\mathfrak{bf};\mathfrak{bf}}
	&\propto 32 + \frac{\pi^4 R}{4},
	\\
	\mathcal{E}_{\mathfrak{fb};\mathfrak{bf}}
	=
	\mathcal{E}_{\mathfrak{bf};\mathfrak{fb}}
	&\propto -32 + \frac{\pi^4 R}{4},
\end{align}
yielding upon normalization:
\begin{equation}
	\label{eq:rhoayhemi}
	\rho	=
		\mqty(0 & 0 & 0 & 0\\
				 0 & \dfrac 12 & \dfrac12 - \dfrac{128}{128 + \pi^4 R} & 0 \\
				 0 & \dfrac12 - \dfrac{128}{128 + \pi^4 R} & \dfrac12& 0\\
				 0 & 0 & 0 & 0).
\end{equation}
The density matrix in \eq{eq:rhoayhemi} admits the following Fano decomposition
\begin{equation}
	\vec B^+ = \vec B^- = \vec 0, \quad C = \mqty(1 - \frac{256}{128+\pi^4 R} & 0 & 0 \\ 0 & 1 - \frac{256}{128+\pi^4 R} & 0 \\ 0 & 0 & -1 ),
\end{equation}
and the quantum information observables are expressed as
\begin{align}
	\tr(\rho^2) &= \frac{128^2+\pi^8R^2}{\qty(128+\pi^4R)^2},
	\\
	\mathcal{C}[\rho] &= -1 +\frac{256}{128 + \pi^4 R},
	\\
	\mathfrak{m}_{12} & = 1+ \qty(\frac{128-\pi^4 R}{128+\pi^4 R})^{2},
\end{align}
which are plotted in Fig.~\ref{fig:plotphiEnt} as functions of $R$. At the Belle~II energy, $R\simeq0.113$, they give $\tr(\rho^2)\simeq0.85$, $\mathcal{C}[\rho]\simeq0.84$ and $\mathfrak{m}_{12}\simeq1.71$. The weight of the coarse-grained state is $w=3\,(128+\pi^4R)/[64\pi^2(2+R)]$, equal to $0.31$ at the same energy.

In the $m_\tau\to0$ limit
\begin{equation}
	\label{eq:rhoy_massless}
	\rho
	\xrightarrow{R\to0 \,(s\gg m_\tau^2)}
				 \mqty(0 & 0 & 0 & 0\\
				 0 & \frac 12 & -\frac12 & 0 \\
				 0 &  -\frac12 & \frac12& 0\\
				 0 & 0 & 0 & 0) = \dyad{\Psi^-}
\end{equation}
where
\begin{equation}
	\ket{\Psi^-} = \frac{1}{\sqrt 2} \qty(\ket{\mathfrak{f};\mathfrak{b}}-\ket{\mathfrak{b};\mathfrak{f}})
\end{equation}
is a Bell state. Hence, in this limit we have:
\begin{equation}
	\tr(\rho^2)=1, \quad \mathcal{C}[\rho]=1, \quad \mathfrak{m}_{12} = 2,
\end{equation}
that is, the state is pure, maximally entangled and Bell nonlocal.
We emphasize that the Bell nonlocality observed in the momentum states is a property of the state as predicted by the Standard Model~\cite{Fabbrichesi:2025aqp}. Our hemisphere-projector observables commute and therefore cannot provide the incompatible settings necessary to experimentally test a Bell/CHSH inequality~\cite{Bell:1964,Clauser:1969ny}.

\section{Spin and momentum}
\label{sec:spin_momentum}

In the previous section the spin states were traced out. We now reinstate them, and construct the joint state of the hemisphere-discretized
momenta and the helicities.

\subsection{Joint states}

Each $\tau$ carries a hemisphere qubit $h\in\{\mathfrak{f},\mathfrak{b}\}$ and a spin qubit $s\in\{-,+\}$. We order the four qubits as
$(h_1,s_1;h_2,s_2)$, with particle 1 the $\tau^+$ and particle 2 the $\tau^-$. For each helicity configuration $(r_1r_2)$ of the
initial positron and electron, the hemisphere-discretized state is
\be
\ket{\psi_{r_1r_2}}=\sum_{h}\sum_{s_1s_2}A^{r_1r_2}_{s_1s_2}(h)\,\ket{h;\bar h}\otimes\ket{s_1s_2},\qquad
A^{r_1r_2}_{s_1s_2}(h)=\int_h \dd\Omega\;\mathcal M^{r_1r_2}_{s_1s_2}(\Omega),
\label{eq:joint_def}
\ee
which is the construction of Sec.~\ref{sec:momentum} without the partial trace over the helicities. For unpolarized beams the two
configurations are mixed with equal weights. Tracing over $s_1,s_2$ gives back the momentum states of the previous section. We use
the notation
\be
\ket{\Psi^\pm}_p=\frac{\ket{\mathfrak{f};\mathfrak{b}}\pm\ket{\mathfrak{b};\mathfrak{f}}}{\sqrt2},\qquad
\ket{\Psi^\pm}_s=\frac{\ket{-+}\pm\ket{+-}}{\sqrt2},
\ee
for the momentum and helicity Bell states. The momentum states $\ket{\Psi^\pm}_p$ are the $\ket{\Psi^\pm}$ of Sec.~\ref{sec:momentum}.

Momentum conservation keeps only $\ket{\mathfrak{f};\mathfrak{b}}$ and $\ket{\mathfrak{b};\mathfrak{f}}$, so that, when they are considered together as in the separation of the momenta from the spins, the two hemisphere qubits act as a
single momentum qubit, which records whether the $\tau^+$ went forward or backward. In this qubit, $\ket{\Psi^+}_p$ is the symmetric
superposition of ``$\tau^+$ forward'' and ``$\tau^+$ backward'', and $\ket{\Psi^-}_p$ the antisymmetric one. The joint state is
therefore a state of one momentum qubit and two helicity qubits.

Momentum conservation does not remove any of the four qubits: it restricts the two hemisphere qubits to the two-dimensional subspace spanned by $\ket{\mathfrak{f};\mathfrak{b}}$ and $\ket{\mathfrak{b};\mathfrak{f}}$, and whether this subspace counts as one qubit or as two depends on the bipartition. In the separation $p|s$ both hemisphere qubits lie on the same side, which reduces to the single momentum qubit described above: the bipartition is of the $2\times4$ type, and the negativity is at most $1/2$. In the separation $\tau^+|\tau^-$ they lie on opposite sides, and cannot be merged without destroying the bipartition itself: each $\tau$ is a four-dimensional system, made of its hemisphere qubit and its helicity qubit, and the negativity can reach $3/2$. For the same reason, the momentum state of Sec.~\ref{sec:momentum} is a genuine state of two qubits, one for each $\tau$, even though its support is two-dimensional, as for the Bell states $\ket{\Psi^\pm}_p$. 

\subsection{Defining the spin}
\label{sec:spin_def}

A spin label attached to an integral over directions, as in \eq{eq:joint_def}, requires identifying the spin states at different
momenta. For the Hilbert space of each particle to factorize as momentum $\otimes$ spin, the spin operators must act within the space
at fixed momentum, $[S_i,P_j]=0$, and satisfy the $su(2)$ algebra. This condition is necessary but does not single out a unique spin description.
Any operator obtained from the canonical spin by a rotation that depends only on the momentum, $S'_a=\sum_i e_a^i(\hat P)\,S_i$, also
commutes with $\vec P$, since the triad vectors $\vec e_a(\hat P)$ are functions of $\vec P$. The helicity triad, with $\hat k=\hat p$ and
the two other axes obtained from the $x$ and $y$ axes by the rotation that takes the beam direction into $\hat p$, is of this form, and indeed the helicity operator itself commutes with the momentum:
$[\vec J\cdot\hat P,P_j]=i\epsilon_{ijk}\hat P_iP_k=0$. The admissible momentum $\otimes$ spin decompositions thus form a family related
by momentum-dependent $SU(2)$ rotations.

The canonical description, in which the spin is quantized along a fixed axis for all momenta, is selected if two further requirements are imposed. The first is rotation covariance:
$\vec S$ must be a vector operator, $[J_i,S_j]=i\epsilon_{ijk}S_k$, so that $\vec L=\vec J-\vec S$ is the orbital angular momentum
$\vec X\times\vec P$ of a position operator, as in the Newton--Wigner and Pryce constructions~\cite{Newton:1949cq,Pryce:1948pf}. The spin defined by the helicity triad is not a vector operator: its transverse axes are fixed relative to the beam, and the rotation
$R(\phi,\theta,-\phi)$ that defines them is singular at $\theta=\pi$ and produces the $e^{\mp i\phi}$ and $e^{\mp 2i\phi}$ phases of the helicity amplitudes.
The third Euler angle is itself a convention, and the coarse-grained state depends on it. For instance, if the helicity states are defined via the rotation $R(\phi,\theta,0)$, all the amplitudes carry
the common phase $e^{\mp i\phi}$, the projection onto the hemisphere states along the beam vanishes, and the one along $y$ gives
$\ket{\Psi^-}_p$ at all energies. The second
requirement is parity, which removes the residual freedom of a rotation about $\hat p$ itself, since a term $\hat p\times\vec S$ has the
wrong parity for a spin. Notice that the rotation by $\pi$ about $\hat p$, which parity allows, is excluded by continuity at $\vec p\to0$. Even the canonical decomposition refers to a frame: under boosts the spins undergo Wigner rotations that depend on the momentum, so the decomposition, and the entanglement across it, depend on the frame~\cite{Peres:2002wx}. Our results refer to the $e^+e^-$ center-of-mass frame.

Momentum and spin can be entangled even though the spin operators commute with the momenta. Commutation is what makes them two
separate subsystems, so that the Hilbert space splits as $\mathscr H_{p}\otimes\mathscr H_{s}$, and entanglement is a property of the
state across this split. In the same way, $\sigma_z\otimes\mathbb 1$ and $\mathbb 1\otimes\sigma_z$ commute, while
$(\ket{00}+\ket{11})/\sqrt2$ is maximally entangled, and a Stern--Gerlach magnet turns a spin superposition with a single momentum into
the spin--momentum entangled state $\ket{\uparrow}\ket{p_{\uparrow}}+\ket{\downarrow}\ket{p_{\downarrow}}$. Different commuting triads,
however, define different splits. Two of them are related by a momentum-controlled rotation,
\be
U=\sum_p\dyad{p}\otimes R(p),
\ee
which acts across the momentum--spin split as a controlled gate and can create or remove entanglement between momentum and spin.
The amount of momentum--spin entanglement, therefore, depends on the choice of triad as well as on the state. We give the results in the
helicity basis, used in the paper so far, and in the canonical basis.

\subsection{Helicity basis}

The amplitudes fall into two families: helicity-conserving amplitudes, with opposite helicities ($-+$ or $+-$), of order one and
carrying the azimuthal phases $e^{\mp i\phi}$; and helicity-flip amplitudes, with equal helicities ($--$ or $++$), suppressed by
$2m_\tau/\sqrt s=\sqrt R$ with $R$ defined in \eq{eq:rho_mat_el_p}, of which those that survive the integration do not depend on $\phi$. The hemisphere integration selects them
by their azimuthal mode:
\begin{center}
\setlength{\tabcolsep}{10pt}\renewcommand{\arraystretch}{1.3}
\begin{tabular}{lll}
\hline
axis & helicity-conserving ($e^{\mp i\phi}$) & helicity-flip ($\phi$-independent)\\[3pt]
\hline
\noalign{\vskip 3pt}
$y$ & opposite signs in $\mathfrak{f}$ and $\mathfrak{b}$: $\ket{\Psi^-}_p$ & equal in $\mathfrak{f}$ and $\mathfrak{b}$: $\ket{\Psi^+}_p$\\
$z$ & integrate to zero & survive: $\ket{\Psi^+}_p$\\
\hline
\end{tabular}
\end{center}

\paragraph{$y$ hemispheres.}
The azimuthal integrals are $\int_{\mathfrak f} e^{\mp i\phi}\dd\phi=\mp2i$, $\int_{\mathfrak b} e^{\mp i\phi}\dd\phi=\pm2i$,
$\int e^{\pm2i\phi}\dd\phi=0$ and $\int\dd\phi=\pi$ over either hemisphere. The ones involving the polar angle are $\int(\cos\theta\pm1)\,\dd\cos\theta=\pm2$
and $\int\sin\theta\,\dd\cos\theta=\pi/2$. The opposite weights given to $\ket{-+}$ and $\ket{+-}$ pair $\ket{\Psi^-}_p$ with
$\ket{\Psi^-}_s$, and from \eq{eq:joint_def}
\begin{align}
\ket{\psi_{-+}}&=-8\,\ket{\Psi^-}_p\ket{\Psi^-}_s+i\,\frac{\pi^2\sqrt R}{\sqrt2}\,\ket{\Psi^+}_p\ket{--},\nn\\
\ket{\psi_{+-}}&=-8\,\ket{\Psi^-}_p\ket{\Psi^-}_s-i\,\frac{\pi^2\sqrt R}{\sqrt2}\,\ket{\Psi^+}_p\ket{++}.
\label{eq:joint_y}
\end{align}
With $\ket X=\ket{\Psi^-}_p\ket{\Psi^-}_s$, $\ket{Y_\pm}=\ket{\Psi^+}_p\ket{\pm\pm}$
and $N=128+\pi^4R$, the state for unpolarized beams is
\be
\rho^{(y)}=\frac1N\Big[128\,\dyad{X}+\frac{\pi^4R}{2}\big(\dyad{Y_-}+\dyad{Y_+}\big)
+4\sqrt2\,i\,\pi^2\sqrt R\,\big(\ketbra{X}{Y_-}-\ketbra{X}{Y_+}\big)+\text{h.c.}\Big],
\label{eq:rhoy_joint}
\ee
an equal mixture of two pure states with overlap $128/N$, with eigenvalues $\frac12(1\pm128/N)$ and purity
$\frac12[1+(128/N)^2]$. Its reduced states are
\begin{align}
\rho_p&=\tr_s\rho^{(y)}=\frac1N\Big[128\,\dyad{\Psi^-}_p+\pi^4R\,\dyad{\Psi^+}_p\Big],\nn\\
\rho_s&=\tr_p\rho^{(y)}=\frac1N\Big[128\,\dyad{\Psi^-}_s+\frac{\pi^4R}{2}\big(\dyad{--}+\dyad{++}\big)\Big].
\end{align}
The momentum state is that of \eq{eq:rhoayhemi}, and the two reduced states have the same concurrence,
$\mathcal C[\rho_p]=\mathcal C[\rho_s]=(128-\pi^4R)/N$.

In the massless limit only the first term of \eq{eq:rhoy_joint} survives, and both beam helicities give the
pure state $\ket{\Psi^-}_p\otimes\ket{\Psi^-}_s$: the $\tau^+$ and $\tau^-$ share one ebit in momentum (encoding which hemisphere) and one in
helicity, independent of each other. In quantum optics, a pair entangled in two degrees of freedom at once, with no correlation between
them, is called hyperentangled~\cite{Kwiat:1997}. For $R>0$ each beam helicity gives a coherent superposition of a state antisymmetric in the momenta
with opposite helicities and a state symmetric in the momenta with equal helicities. Projecting the momentum onto $\ket{\Psi^+}_p$ leaves
equal helicities, projecting it onto $\ket{\Psi^-}_p$ leaves $\ket{\Psi^-}_s$: momentum and spin are entangled with each other. Each pure
component has Schmidt rank two between momenta and spins, and its concurrence is
\be
\mathcal C(p|s)=\frac{16\sqrt2\,\pi^2\sqrt R}{128+\pi^4R},
\ee
equal to $0.54$ at Belle~II. The two beam helicities put the mass term on orthogonal spin states, $\ket{--}$ and $\ket{++}$, so their
mixture keeps the common first term but loses part of the coherence.

This also explains why the momentum state of \eq{eq:rhoayhemi} is mixed, with $\mathcal C=0.84$ rather than $1$ at Belle~II. In each
pure component $\ket{\Psi^-}_p$ is paired with $\ket{\Psi^-}_s$ and $\ket{\Psi^+}_p$ with $\ket{\pm\pm}$. These spin partners are
orthogonal, so the trace over the spins removes the coherence between $\ket{\Psi^-}_p$ and $\ket{\Psi^+}_p$. The same mechanism, with
the roles exchanged, makes the spin state mixed. The two reduced states have the same concurrence only for unpolarized beams: for a
single beam helicity $\mathcal C[\rho_s]=128/N$, equal to $0.92$ at Belle~II. Tracing out the spins is therefore not a neutral simplification: the loss of purity of the momentum state measures how strongly the momenta are entangled with the discarded spins.

\begin{figure}[h!]
\centering
\includegraphics[width=0.6\linewidth]{"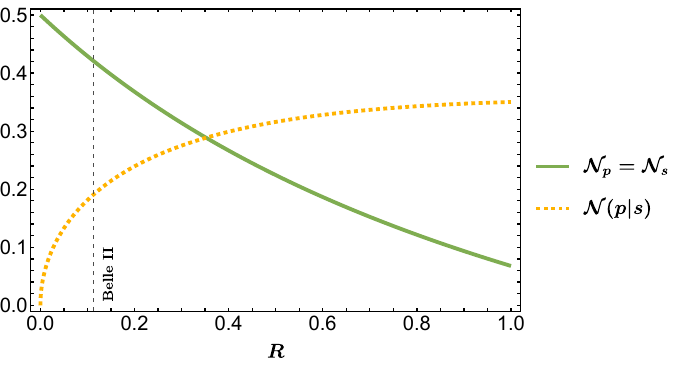"}
\caption{Negativity of the momentum state $\mathcal N_p$ and of the spin state $\mathcal N_s$ (solid; the two coincide) and
negativity between momenta and spins $\mathcal N(p|s)$ (dashed), as functions of $R=4m_\tau^2/s$, for the $y$ hemispheres in the
helicity basis. The dashed vertical line marks Belle~II. For the $z$ hemispheres, and in the canonical basis, $\mathcal N_p=1/2$ and
$\mathcal N_s=\mathcal N(p|s)=0$ for all $R$.}
\label{fig:negR}
\end{figure}

Table~\ref{tab:joint} gives the entanglement for the main bipartitions, quantified by the negativity $\mathcal N$, and
Fig.~\ref{fig:negR} shows its dependence on $R$. The negativity of the momentum state coincides with that of the spin state,
\be
\mathcal N_p=\mathcal N_s=\frac{128-\pi^4R}{2\,(128+\pi^4R)}=\frac{\mathcal C}{2},
\ee
and both decrease from the maximal value $1/2$ at $R=0$ to $0.07$ as $R\to1$. The momentum--spin negativity grows from zero, overtakes them
at $R\simeq0.35$ and approaches $0.35$ as $R\to1$: increasing $R$ transfers entanglement from within the momentum and the spin sectors
to between them. Each $\tau$ is entangled with the other but not internally: the state of $(h_1,s_1)$ alone is separable.

\begin{table}[h!]
\centering
\setlength{\tabcolsep}{10pt}\renewcommand{\arraystretch}{1.3}
\begin{tabular}{lcccccc}
\hline
& $\tr\rho^2$ & $\mathcal C[\rho_p]$ & $\mathcal C[\rho_s]$ & $\mathcal N(\tau^+|\tau^-)$ & $\mathcal N(p|s)$ & $\mathcal N(h_1|\text{rest})$\\[3pt]
\hline
\noalign{\vskip 3pt}
$z$, any $R$         & 0.500 & 1     & 0     & 0.500 & 0     & 0.500\\
$y$, $R=0$           & 1     & 1     & 1     & 1.500 & 0     & 0.500\\
$y$, $R=0.113$ (Belle~II) & 0.924 & 0.842 & 0.842 & 1.421 & 0.191 & 0.500\\
$y$, $R=0.3$         & 0.831 & 0.628 & 0.628 & 1.314 & 0.275 & 0.500\\
$y$, $R=0.99$        & 0.663 & 0.141 & 0.141 & 1.070 & 0.350 & 0.500\\
\hline
\end{tabular}
\caption{Entanglement of the joint spin--momentum state in the helicity basis. $\mathcal N(\tau^+|\tau^-)$ splits $(h_1,s_1)$ from
$(h_2,s_2)$, with maximal value $3/2$; $\mathcal N(p|s)$ splits $(h_1,h_2)$ from $(s_1,s_2)$.}
\label{tab:joint}
\end{table}

\paragraph{$z$ hemispheres.}
Only the helicity-flip terms survive, and
\be
\rho^{(z)}=\dyad{\Psi^+}_p\otimes\frac12\big(\dyad{--}+\dyad{++}\big),
\ee
the product of a Bell state in momentum and a classically correlated, separable spin state. The whole state comes from the amplitudes
of order $\sqrt R$.

\subsection{Canonical basis}

With the spin quantized along the beam for all momenta we find, numerically and for both the $y$ and the $z$ hemispheres at all
energies,
\be
\rho^{\rm can}=\dyad{\Psi^+}_p\otimes\frac12\big(\dyad{\uparrow\uparrow}+\dyad{\downarrow\downarrow}\big).
\label{eq:can_joint}
\ee
The momentum state is $\ket{\Psi^+}_p$, the spins are classically correlated along the beam, and momentum and spin are not entangled.
The weight of the state is
\be
w=\frac{(2+\sqrt R)^2}{3\,(2+R)},
\label{eq:weight_can}
\ee
equal to $0.86$ at Belle~II.

This follows from parity. With a momentum-independent spin basis the amplitude factorizes into orbital and spin parts,
$\mathcal M=\sum c_{Lm_LSm_S}Y_{Lm_L}(\Omega)\ket{S\,m_S}$. The $\tau^+\tau^-$ pair produced by a virtual photon has $J^{PC}=1^{--}$, hence
$S=1$ and $L=0,2$. The hemisphere combinations are
\be
A_{\mathfrak f}+A_{\mathfrak b}=\int\dd\Omega\,\mathcal M,\qquad A_{\mathfrak f}-A_{\mathfrak b}=\int\dd\Omega\,\operatorname{sgn}(\hat n\cdot\hat p)\,\mathcal M,
\ee
and $\operatorname{sgn}(\hat n\cdot\hat p)$ is odd under $\vec p\to-\vec p$, so it contains only odd $L$. Therefore
$A_{\mathfrak f}-A_{\mathfrak b}=0$ and the momentum state is $\ket{\Psi^+}_p$ for any axis. The sum projects onto the $S$ wave, in which the
spins carry the $J_z=\pm1$ of the virtual photon along the beam, that is, $\ket{\uparrow\uparrow}$ or $\ket{\downarrow\downarrow}$ for the
two beam helicities. The weight of \eq{eq:weight_can} is the fraction of the rate in the $S$ wave. The remaining part is the $D$ wave, in
which the momenta are entangled with the spins. This part is discarded by the projection onto the hemisphere states, so that the absence of
momentum--spin entanglement in \eq{eq:can_joint} holds only for the part of the state that is kept.

The difference between \eq{eq:rhoy_joint} and \eq{eq:can_joint} is due to the triad dependence discussed in
Sec.~\ref{sec:spin_def}. In the canonical basis, orbital and spin angular momentum are separated, and the hemisphere projection reads as
a partial-wave projection. In the helicity basis, part of the orbital angular momentum, carried by the $e^{\pm i\phi}$ phases, is stored
in the helicity labels, and the $\ket{\Psi^-}_p\ket{\Psi^-}_s$ structure and the momentum--spin entanglement are partly a consequence of
this bookkeeping. Both bases give the same cross sections and the same spin correlations at fixed momentum.

\section{Pseudo-spin operators}
\label{sec:pseudospin}

The hemisphere states of Sec.~\ref{sec:momentum} keep only the part of the produced state that lies in two partial waves, and the diagonal elements of the coarse-grained state are generally not the hemisphere probabilities that a detector measures. We now discretize the momenta in a different way, which keeps the whole state and has the measured probabilities on the diagonal of the density matrix. We first define the operators, then give the momentum state and the joint state of momenta and spins, and finally compare the two discretizations. In Sec.~\ref{sec:measurements} we use these operators to discuss what can be measured.

\subsection{Definition}

For each particle we define the pseudo-spin operators~\cite{Chen:2002elr,Gour:2003wsm}
\be
\sigma_z=\Lambda_{\mathfrak f}-\Lambda_{\mathfrak b},\qquad
\sigma_x=\Pi,\qquad
\sigma_y=i\,\sigma_x\sigma_z,
\label{eq:pseudospin}
\ee
where $\Lambda_h$ are the hemisphere projectors of \eq{eq:h_proj1p} and $\Pi$ is the inversion of the momentum at fixed spin label, $\Pi\ket{p,\,s}=\ket{-p,\,s}$. In other words, $\sigma_z$ is the sign of $\vec p\cdot\hat n$ and $\sigma_x$ exchanges the two hemispheres. Since $\Pi^2=\mathbb 1$ and $\Pi\,\Lambda_{\mathfrak f}\,\Pi=\Lambda_{\mathfrak b}$, the three operators square to the identity and satisfy the $su(2)$ algebra. They are the momentum-space analogue of the configuration-space operators of Ref.~\cite{Gour:2003wsm}, with $\sigma_z$ and $\sigma_x$ in place of their $\Pi_x$ and $\Pi_z$, and with the inversion of the momentum in place of the reflection of the coordinate.

The existence of these operators means that the Hilbert space of each particle factorizes into a qubit and a remainder,
\be
\ket{p,\,s}=\ket{h}\otimes\ket{\omega,\,s},
\ee
where $h$ is the hemisphere of $\vec p$ and $\omega$ labels the pair of opposite momenta $\{\vec p,-\vec p\}$. We define the discretized state as the partial trace over the remainders of the two particles. The partial trace is a local operation, so that entanglement of the qubits implies entanglement of the momenta, as for the hemisphere states. Differently from the map of Sec.~\ref{sec:momentum}, it preserves the trace: the whole state is kept, and the diagonal elements of the density matrix are the hemisphere probabilities $\tr[\Lambda_h\rho]$.

For the back-to-back pair, with $\mathcal M_{s_1s_2}(\Omega)$ the amplitude for a $\tau^+$ emitted in the direction $\Omega$, the matrix elements on $\ket{\mathfrak f;\,\mathfrak b}$ and $\ket{\mathfrak b;\,\mathfrak f}$ are
\begin{align}
\mathcal E_{\mathfrak{fb};\mathfrak{fb}}&\propto\int\limits_{\mathfrak f}\dd\Omega\,\sum_{s_1s_2}\abs{\mathcal M_{s_1s_2}(\Omega)}^2,\qquad
\mathcal E_{\mathfrak{bf};\mathfrak{bf}}\propto\int\limits_{\mathfrak b}\dd\Omega\,\sum_{s_1s_2}\abs{\mathcal M_{s_1s_2}(\Omega)}^2,\nn\\
\mathcal E_{\mathfrak{fb};\mathfrak{bf}}&\propto\int\limits_{\mathfrak f}\dd\Omega\,\sum_{s_1s_2}\mathcal M_{s_1s_2}(\Omega)\,\mathcal M^*_{s_1s_2}(-\Omega),
\label{eq:ps_elements}
\end{align}
to be averaged over the beam helicities for unpolarized beams. The diagonal elements are the fractions of events with the $\tau^+$ in either hemisphere. The off-diagonal element is the coherence between each direction and the opposite one. In \eq{eq:rho_mat_el_p}, instead, each direction of one hemisphere interferes with every direction of the other.

The inversion $\Pi$ acts at fixed spin label, so that the result depends, once again, on how the spin is defined at different momenta. In the canonical basis $\Pi$ is the parity operator, up to the intrinsic parity of the particle, and it commutes with the spin. In the helicity basis it is not, because parity reverses the helicity.

\subsection{Momentum state}

In the helicity basis the integrands of \eq{eq:ps_elements} follow from \eq{eq:rho_mat_el_p}. For $\Omega'=\Omega$ the integrand is $2\,(1+\cos^2\theta+R\sin^2\theta)$, proportional to the differential cross section. For $\Omega'=-\Omega$, that is, $\theta'=\pi-\theta$ and $\phi'=\phi+\pi$, it is $-2\,(1-R)\sin^2\theta$. Both are even under the inversion, so that their integrals over a hemisphere are half of those over the full solid angle, whatever the axis $\hat n$. The state is therefore the same for all hemisphere choices,
\begin{equation}
	\label{eq:ps_hel}
	\rho = \frac12
		\mqty(0 & 0 & 0 & 0\\
				 0 & 1 & -\kappa & 0 \\
				 0 & -\kappa & 1 & 0\\
				 0 & 0 & 0 & 0)
	= \frac{3\,\dyad{\Psi^-}+(1+2R)\,\dyad{\Psi^+}}{2\,(2+R)},
	\qquad
	\kappa=\frac{1-R}{2+R},
\end{equation}
with the Fano coefficients $\vec B^+=\vec B^-=\vec 0$ and $C=\text{diag}(-\kappa,\,-\kappa,\,-1)$, and
\begin{equation}
	\tr(\rho^2)=\frac{1+\kappa^2}{2}, \quad \mathcal{C}[\rho]=\kappa, \quad \mathfrak{m}_{12} = 1+\kappa^2.
\end{equation}
At the Belle~II energy and with the phase convention of Sec.~\ref{sec:momentum}, $\kappa\simeq0.42$, which gives $\tr(\rho^2)\simeq0.59$, $\mathcal{C}[\rho]\simeq0.42$ and $\mathfrak{m}_{12}\simeq1.18$. The state is an entangled and nonlocal mixture of the two Bell states. It has $\kappa=1/2$ in the massless limit and reduces to the classical state of \eq{eq:rhoadiag_ev} at threshold, $R\to1$. The value of $\kappa$ depends on the phase convention of the helicity states, through the inversion $\Pi$: with the rotation $R(\phi,\theta,0)$ discussed in Sec.~\ref{sec:spin_def}, the equal-helicity amplitudes change sign under the inversion, the integrand for $\Omega'=-\Omega$ becomes $-2\,(1+R)\sin^2\theta$, and $\kappa=(1+R)/(2+R)$, equal to $0.53$ at Belle~II. The two conventions agree in the massless limit, where $\kappa=1/2$, and differ most at threshold, where $\kappa$ is $0$ and $2/3$, respectively. The canonical result below does not depend on them.

In the canonical basis parity gives $\mathcal M_{s_1s_2}(-\Omega)=\mathcal M_{s_1s_2}(\Omega)$, because the pair is produced with even orbital angular momentum. The off-diagonal element in \eq{eq:ps_elements} then equals the diagonal ones, and
\begin{equation}
	\label{eq:ps_can}
	\rho = \dyad{\Psi^+}
\end{equation}
for any axis $\hat n$ and at any energy. The state is pure, maximally entangled and Bell nonlocal, and it is the whole state, not a part of it. In fact, the result follows from two properties of the pair alone. Momentum conservation gives $\sigma_z\otimes\sigma_z=-1$, since the two particles are found in opposite hemispheres, and parity gives $\sigma_x\otimes\sigma_x=+1$; the only state with these two eigenvalues is $\ket{\Psi^+}$. The result relies on parity and therefore holds for the photon exchange considered here. At the Belle~II energy, $\gamma$--$Z$ interference violates parity and gives a forward--backward asymmetry below one percent. The hemisphere probabilities change at the same order, while the entanglement of the state decreases only at second order in the asymmetry.

The Bell nonlocality of these states can be expressed through the CHSH inequality~\cite{Clauser:1969ny} written with the pseudo-spin operators, as proposed for continuous-variable states in Refs.~\cite{Chen:2002elr,Gour:2003wsm}. For settings $A=\hat a\cdot\vec\sigma$, $A'=\hat a'\cdot\vec\sigma$ on the first particle and $B$, $B'$ on the second, the CHSH combination
\be
\mathcal S=\langle A\otimes B\rangle+\langle A\otimes B'\rangle+\langle A'\otimes B\rangle-\langle A'\otimes B'\rangle
\ee
satisfies $|\mathcal S|\le2$ in any local hidden-variable model, while its maximum over the settings is $\mathcal S_{\rm max}=2\sqrt{\mathfrak m_{12}}$~\cite{Horodecki:1995nsk}. For the Bell state of \eq{eq:ps_can}, $\mathcal S_{\rm max}=2\sqrt2$, reached for instance with $A=\sigma_z$, $A'=\sigma_x$ and $B,B'=(-\sigma_z\pm\sigma_x)/\sqrt2$. For the state of \eq{eq:ps_hel}, $\mathcal S_{\rm max}=2\sqrt{1+\kappa^2}$, equal to $2.17$ at Belle~II, or $2.26$ with the alternative phase convention, above the classical bound for any $\kappa>0$. Every choice of settings that violates the inequality involves $\sigma_x$ or $\sigma_y$, that is, a measurement of whether the particle is in a given superposition of $\vec p$ and $-\vec p$. As discussed in Sec.~\ref{sec:measurements}, such measurements are not available at colliders, and the violation is a property of the predicted state.

\subsection{Joint state of momenta and spins}

Reinstating the spins, the state for the beam helicities $(r_1r_2)$ is
\be
\rho_{r_1r_2}\propto\int\limits_{\mathfrak f}\dd\Omega\;\dyad{\varphi_{r_1r_2}(\Omega)},\qquad
\ket{\varphi_{r_1r_2}(\Omega)}=\sum_{s_1s_2}\Big[\mathcal M^{r_1r_2}_{s_1s_2}(\Omega)\,\ket{\mathfrak f;\,\mathfrak b}+\mathcal M^{r_1r_2}_{s_1s_2}(-\Omega)\,\ket{\mathfrak b;\,\mathfrak f}\Big]\otimes\ket{s_1s_2},
\label{eq:ps_joint_def}
\ee
which replaces \eq{eq:joint_def}. The state is mixed even for a single beam helicity, because the label $\omega$ of the direction has been traced out.

In the helicity basis the inversion exchanges the two helicity labels, $\mathcal M_{s_1s_2}(-\Omega)=\mathcal M_{s_2s_1}(\Omega)$. The part of the amplitude that is symmetric in the two labels is then paired with $\ket{\Psi^+}_p$, and the antisymmetric one with $\ket{\Psi^-}_p$. For unpolarized beams and with the phase convention of Sec.~\ref{sec:momentum} we find, for both the $y$ and the $z$ hemispheres,
\be
\rho^{\rm hel}=\frac{1}{2\,(2+R)}\Big[3\,\dyad{\Psi^-}_p\otimes\dyad{\Psi^-}_s
+\dyad{\Psi^+}_p\otimes\big(\dyad{\Psi^+}_s+R\,\dyad{--}+R\,\dyad{++}\big)\Big].
\label{eq:ps_joint_hel}
\ee
This is a mixture of four orthogonal products of a momentum state and a spin state: momenta and spins are classically correlated but not entangled, $\mathcal N(p|s)=0$. The reduced momentum state is that of \eq{eq:ps_hel}, the reduced spin state has the same concurrence $\kappa$, and
\be
\tr\rho^2=\frac{5+R^2}{2\,(2+R)^2},\qquad \mathcal N(\tau^+|\tau^-)=\frac12+\kappa.
\ee
In the massless limit the state is $\frac34\,\dyad{\Psi^-}_p\otimes\dyad{\Psi^-}_s+\frac14\,\dyad{\Psi^+}_p\otimes\dyad{\Psi^+}_s$, with purity $5/8$: the hyperentangled state found in Sec.~\ref{sec:spin_momentum} is specific to the hemisphere states. Notice that the entanglement between momenta and spins is not absent from the amplitudes. For a single beam helicity the terms that mix the two momentum Bell states do not vanish, and $\mathcal N(p|s)=0.35$ for the $z$ and $0.12$ for the $y$ hemispheres at Belle~II. These terms, however, cancel between the two beam helicities.

In the canonical basis $\mathcal M_{s_1s_2}(-\Omega)=\mathcal M_{s_1s_2}(\Omega)$ and the state factorizes,
\be
\rho^{\rm can}=\dyad{\Psi^+}_p\otimes\bar\rho_s,\qquad
\bar\rho_s=a\,\big(\dyad{\uparrow\uparrow}+\dyad{\downarrow\downarrow}\big)+b\,\dyad{\Psi^+}_s,
\label{eq:ps_joint_can}
\ee
for any axis $\hat n$, with
\be
a=\frac{9+2\sqrt R+4R}{10\,(2+R)},\qquad b=\frac{(1-\sqrt R)^2}{5\,(2+R)},
\ee
and $\ket{\Psi^+}_s=(\ket{\uparrow\downarrow}+\ket{\downarrow\uparrow})/\sqrt2$. Here $\bar\rho_s$ is the spin state of the pair averaged over the directions. With respect to \eq{eq:can_joint}, the $D$ wave now contributes, through the term proportional to $b$, equal to $0.04$ at Belle~II. The spin state is separable, and momenta and spins are not entangled. Table~\ref{tab:pseudospin} gives the entanglement of the two joint states for the bipartitions of Table~\ref{tab:joint}.

\begin{table}[h!]
\centering
\setlength{\tabcolsep}{10pt}\renewcommand{\arraystretch}{1.3}
\begin{tabular}{lcccccc}
\hline
& $\tr\rho^2$ & $\mathcal C[\rho_p]$ & $\mathcal C[\rho_s]$ & $\mathcal N(\tau^+|\tau^-)$ & $\mathcal N(p|s)$ & $\mathcal N(h_1|\text{rest})$\\[3pt]
\hline
\noalign{\vskip 3pt}
helicity, $R=0$                  & 0.625 & 0.500 & 0.500 & 1.000 & 0 & 0.500\\
helicity, $R=0.113$ (Belle~II)   & 0.561 & 0.420 & 0.420 & 0.920 & 0 & 0.500\\
helicity, $R=0.3$                & 0.481 & 0.304 & 0.304 & 0.804 & 0 & 0.500\\
helicity, $R=0.99$               & 0.334 & 0.003 & 0.003 & 0.503 & 0 & 0.500\\
canonical, $R=0.113$ (Belle~II)  & 0.461 & 1     & 0     & 0.500 & 0 & 0.500\\
\hline
\end{tabular}
\caption{Entanglement of the joint spin--momentum state obtained with the pseudo-spin operators, for unpolarized beams and any choice of
hemispheres. The bipartitions are those of Table~\ref{tab:joint}.}
\label{tab:pseudospin}
\end{table}

\subsection{Comparison of the two discretizations}

The two discretizations we considered probe different properties of the same state. The hemisphere states of Sec.~\ref{sec:momentum} select two partial waves of the pair, and the coarse-grained state gives the coherence between them. The pseudo-spin operators keep all the waves, and the state gives the coherence between each configuration of the pair and the inverted one. In the canonical basis the two agree, and for the same reason: parity forbids the odd waves in the first case and makes the amplitude even under the inversion in the second. The hemisphere states find the Bell state in the $S$ wave, a fraction $0.86$ of the rate at Belle~II, while the pseudo-spin operators find it in the whole state. In the helicity basis the two differ, in the momentum state as well as in its relation with the spins. Table~\ref{tab:comparison} summarizes the results.

\begin{table}[h!]
\centering
\setlength{\tabcolsep}{10pt}\renewcommand{\arraystretch}{1.3}
\begin{tabular}{ll@{\qquad}l@{\qquad}l}
\hline
spin basis & & hemisphere states & pseudo-spin operators\\[3pt]
\hline
\noalign{\vskip 3pt}
canonical & momentum state    & $\ket{\Psi^+}$, $w=0.86$ & $\ket{\Psi^+}$\\
          & $\mathcal N(p|s)$ & $0$ & $0$\\
helicity  & momentum state    & $z$: $\ket{\Psi^+}$, $w=0.025$;\quad $y$: $\mathcal C=0.84$, $w=0.31$ & $\mathcal C=0.42$ (0.53)\\
          & $\mathcal N(p|s)$ & $z$: $0$;\quad $y$: $0.19$ & $0$\\
\hline
\end{tabular}
\caption{Momentum state and negativity between momenta and spins in the two discretizations, at the Belle~II energy and for unpolarized
beams. The weight $w$ is the fraction of the produced state kept by the hemisphere states; the pseudo-spin operators keep the whole
state. In the helicity basis the values refer to the phase convention of Sec.~\ref{sec:momentum}. If the helicity spinors are defined through the rotation $R(\phi,\theta,0)$, the hemisphere states give $\ket{\Psi^-}$ with $w=0.31$ and $\mathcal N(p|s)=0$ for $y$ and zero weight for $z$, and the pseudo-spin operators give the concurrence in parentheses, with $\mathcal N(p|s)=0.08$ for $z$ and $0$ for $y$.}
\label{tab:comparison}
\end{table}

\section{Experimental measurements}
\label{sec:measurements}

What a collider experiment measures is the direction of each particle and, through the angular distributions of the decay products, its spin. The hemisphere in which each particle is found gives the probabilities $p_h=\tr[\Lambda_h\rho]$ of \eq{eq:h_proj1p}, obtained from the differential cross section integrated over each hemisphere. For $\tau^+\tau^-$ production they equal $p_{\mathfrak{f}\mathfrak{b}}=p_{\mathfrak{b}\mathfrak{f}}=1/2$ for any axis, as a consequence of the symmetry of the cross section. With $N$ events the statistical uncertainty on $p_{\mathfrak{f}\mathfrak{b}}$ is $1/(2\sqrt N)$, about $5\times10^{-4}$ for a sample of $N\simeq10^{6}$ events, a small fraction of the $\tau$ pairs already collected by Belle~II. We discuss what these measurements can establish for the two discretizations in turn.

\subsection{Pseudo-spin operators}

In the setup of Sec.~\ref{sec:pseudospin}, the hemisphere in which a particle is found is the eigenvalue of $\sigma_z$, and the probabilities $p_h$ are the diagonal elements of the pseudo-spin states of \eq{eq:ps_hel} and \eq{eq:ps_can}. The entropy of the reduced single-particle state, which we discuss for the hemisphere states in Sec.~\ref{sec:meas_hemi}, is $\log2$ also here, since the diagonal elements of the pseudo-spin states are exactly the measured probabilities $p_h=1/2$. Also in this case, it is not experimentally possible to distinguish the entangled states from the classical mixture of \eq{eq:rhoadiag_ev}. In the helicity basis, the state of \eq{eq:ps_hel} is not even pure: $\tr(\rho^2)\simeq0.59$ at the Belle~II energy. What distinguishes these states is the correlation $\langle\sigma_x\otimes\sigma_x\rangle$, equal to $+1$ for the Bell state of \eq{eq:ps_can}, to $-\kappa$ for the state of \eq{eq:ps_hel} and to zero for the classical mixture. A measurement of $\sigma_x$ would require the interference of each direction with the opposite one, which no detector can provide. The pseudo-spin operators of each particle do not commute with each other, but those that can be measured do: all the observables accessible at a collider are functions of the momenta, together with the spin observables at fixed momentum, and are therefore diagonal in the momentum basis. In pseudo-spin language they include $\sigma_z$, and any finer partition of the directions, but never $\sigma_x=\Pi$ or $\sigma_y$. Neither the free propagation, which conserves both $\sigma_z$ and $\Pi$, nor a magnetic field, which maps momentum eigenstates onto momentum eigenstates, turns one into the other. Since the measurable functions of the momenta all commute, they admit a joint probability distribution, and by Fine's theorem~\cite{Fine:1982lwu} a local hidden-variable model reproduces all the momentum data that a collider can provide. No Bell inequality for the momenta can then be violated by these data, whatever the state: the violations of Eqs.~\eqref{eq:ps_hel} and \eqref{eq:ps_can}, with $\mathfrak m_{12}=1+\kappa^2$ and $2$, require $\langle\sigma_x\otimes\sigma_x\rangle$ and $\langle\sigma_y\otimes\sigma_y\rangle$. The entropy of the reduced state does not change this: the single-particle pseudo-spin state is $\frac12\mathbb 1$ for the states of Eqs.~\eqref{eq:rhoadiag_ev}, \eqref{eq:ps_hel} and \eqref{eq:ps_can} alike, so that the measured entropy is $\log2$ whatever the value of $\kappa$. This figure is the entanglement of the state only under the assumption that the state is pure, which cannot be verified without the off-diagonal elements. Pseudo-spin Bell tests have been proposed in quantum optics for continuous variables of light, such as the quadratures of two-mode squeezed states~\cite{Chen:2002elr,Gour:2003wsm}, where parity-type observables can be measured through displacements and photon counting~\cite{Banaszek:1998wcc}. Unfortunately, collider experiments offer no such measurement. The discretized momentum states are therefore predictions of the $S$-matrix formalism, and the hemisphere counts provide a consistency check of them, not a reconstruction.

The same limitation applies to the entanglement between momenta and spins. A collider
experiment can measure the spins in any basis, but it measures the momenta
only through $\sigma_z$: it determines whether the $\tau^+$ went forward or backward ($\ket{\mathfrak{f};\mathfrak{b}}$ or
$\ket{\mathfrak{b};\mathfrak{f}}$), never in which superposition ($\ket{\Psi^\pm}_p$). When one side of a bipartition is measured in a single
fixed basis, all the statistics that can be collected are reproduced exactly by the dephased state
\be
D(\rho)=\sum_{h}\dyad{h;\bar h}\,\rho\,\dyad{h;\bar h}=\sum_h p_h\,\dyad{h;\bar h}\otimes\rho^{(h)}_s ,
\label{eq:dephased}
\ee
where $\rho$ is the joint state of \eq{eq:ps_joint_def} and $\rho^{(h)}_s$ is the spin state of the events with the $\tau^+$ in the hemisphere $h$. This
is a classical--quantum state and is therefore separable between momenta and spins. No set of such measurements can witness
momentum--spin entanglement, however well the spins are reconstructed, and with a single measurement setting on the momentum side EPR
steering is excluded as well. The same holds if the directions are resolved more finely than the two hemispheres: the data are then
reproduced by the state diagonal in the momenta.

For unpolarized beams the states of \eq{eq:ps_joint_hel} and
\eq{eq:ps_joint_can} are already separable across this bipartition, the spin state is the same in the two hemispheres,
$\rho^{(\mathfrak f)}_s=\rho^{(\mathfrak b)}_s$, and $D(\rho)$ is the product of the classical state of \eq{eq:rhoadiag_ev} and the spin
state of the pair. For a single beam helicity, and in the helicity basis, the momenta are entangled with the spins, and it is this
entanglement that dephasing cancels. Only a correlation between the hemisphere and the helicities survives, as $\rho^{(\mathfrak b)}_s$ is
$\rho^{(\mathfrak f)}_s$ with the two helicity labels exchanged. With polarized beams this correlation is accessible to spin tomography
conditioned on the hemisphere, but a classical mixture reproduces it.

\subsection{Hemisphere states}
\label{sec:meas_hemi}

For the hemisphere states of Secs.~\ref{sec:momentum} and~\ref{sec:spin_momentum} the connection with the data is weaker. The diagonal elements of the coarse-grained state are not in general the hemisphere probabilities. The two coincide here, at $1/2$, only because of the symmetry of the cross section. The off-diagonal element is the coherence between the uniform superpositions over the two hemispheres, and a measurement of it would require the interference of all the directions of one hemisphere with all those of the other. Interference of this kind between opposite directions has been observed for pairs of particles at low energies: for photon pairs entangled in momentum, in a two-particle interferometer that violated a Bell inequality~\cite{Rarity:1990zz}, and for pairs of metastable helium atoms with opposite momenta, scattered out of a Bose--Einstein condensate by a moving optical lattice or by the collision of two condensates, whose paths are recombined by Bragg pulses of light acting as mirrors and beam splitters~\cite{Lopes:2015qdr,Dussarrat:2017wwi,Athreya:2025jrg}. The last experiment observed correlations strong enough for the violation of a Bell inequality, but with the same phase on the two sides, so that no inequality was actually tested. These devices rely on diffracting structures of the size of the de~Broglie wavelength and on the coherence of the particles until their paths are recombined, and neither is available for unstable particles with GeV energies. This is the two-path interference required by the hemisphere states. For the pseudo-spin states the corresponding observable is the parity, discussed in the previous subsection. 

For a pure two-particle state, the entanglement is measured by the von Neumann entropy of the reduced single-particle state. For a pure coarse-grained state of the momentum qubits, $\ket\psi=a\ket{\mathfrak f;\mathfrak b}+b\ket{\mathfrak b;\mathfrak f}$, the reduced state is diagonal in the hemisphere basis, with eigenvalues $|a|^2$ and $|b|^2$, and the entropy
\be
\mathscr E[\rho]=-\,p_{\mathfrak f\mathfrak b}\log p_{\mathfrak f\mathfrak b}-p_{\mathfrak b\mathfrak f}\log p_{\mathfrak b\mathfrak f},
\qquad p_{\mathfrak f\mathfrak b}=|a|^2,\quad p_{\mathfrak b\mathfrak f}=|b|^2,
\label{eq:entropy}
\ee
is determined by the diagonal elements alone. For $\tau^+\tau^-$ production, mediated by photon exchange, these coincide with the measured hemisphere probabilities, $p_{\mathfrak f\mathfrak b}=p_{\mathfrak b\mathfrak f}=1/2$, so that for the pure states the entropy can be measured from the event counts:
\be
\mathscr E[\rho]=\log2-2\,(\delta p_h)^2+O(\delta p_h^4),
\ee
the maximal value, as expected for the Bell state of \eq{eq:rhoz}. The entropy is stationary at $p_h=1/2$, so that the statistical uncertainty enters at second order: with $\delta p_h=1/(2\sqrt N)$, the deviation from $\log2$ is of order $1/(2N)$, about $5\times10^{-7}$ for $N\simeq10^{6}$ events. The measured entropy, however, cannot be distinguished from the classical result: the classical mixture of \eq{eq:rhoadiag_ev} has the same single-particle state $\frac12\mathbb 1$ and the same entropy $\log2$, which for a mixed state is not a measure of entanglement, and the purity of the two-particle state cannot be established from the hemisphere probabilities.

For the $y$ hemispheres the coarse-grained state of \eq{eq:rhoayhemi} is not even pure at the Belle~II energy, with purity $\tr(\rho^2)\simeq0.85$, and in Sec.~\ref{sec:spin_momentum} we have traced this mixedness to the entanglement of the momenta with the spins.

Keeping the spins, the entanglement between momenta and spins in \eq{eq:joint_y} is the coherence between the two hemisphere branches,
\be
\ket{\mathfrak{f};\mathfrak{b}}\otimes\frac{-8\ket{\Psi^-}_s\pm i\,b\ket{\mp\mp}}{\sqrt2}\quad\text{and}\quad
\ket{\mathfrak{b};\mathfrak{f}}\otimes\frac{8\ket{\Psi^-}_s\pm i\,b\ket{\mp\mp}}{\sqrt2},\qquad b=\frac{\pi^2\sqrt R}{\sqrt2},
\label{eq:branches}
\ee
and it is exactly this coherence that dephasing removes. The spin states of the two branches differ, but this difference is not
accessible either. They are the spin states that accompany the uniform superposition over each hemisphere, whereas the spin state
measured in a hemisphere is $\rho^{(h)}_s$ of \eq{eq:dephased}, the average of the spin states at fixed direction weighted by the angular
distribution. For unpolarized beams it is the same for $\mathfrak f$ and $\mathfrak b$.

In both discretizations, keeping the spins makes one side of the bipartition fully
measurable, and gives access to genuine spin entanglement. Differently, certifying any entanglement that involves the momenta, between the two momenta or between momenta and
spins, still requires interference between the two hemispheres.

That momentum entanglement escapes measurement does not make it a mere formal curiosity. Both the experimental analyses of spin correlations and the event generators that simulate them assign definite momenta to each $\tau$ lepton and confine the quantum description to the helicities, so that the momenta enter only as classical labels of the events. The states found above contradict this assumption. What the $S$ matrix produces is a coherent superposition of directions, in which the momenta of the two leptons are entangled with each other and, in the helicity basis, also with the spins, by an amount that depends on the discretization and on the definition of the spin of Sec.~\ref{sec:spin_def}. The data look classical because every quantity that a detector can record commutes with the momenta, not because the coherence has disappeared from the state. This is the same lesson as Mott's analysis~\cite{Mott1929}, in which straight tracks arise from a spherical wave without any classical ingredient. The spin entanglement measured at colliders should therefore be regarded as a feature of a larger quantum state, in which the momenta are quantum variables in their own right.

\section{Outlook}

We have shown that, by a suitable discretization procedure, the entanglement and nonlocality among the momenta of $\tau$-lepton pairs produced at collider experiments can be characterized with the tools developed for qubits. The $S$-matrix formalism predicts that the momenta are in an entangled state before being measured, which cannot be thought of as a mixture weighted by classical probabilities.

Tracking the spins as well gives a fuller picture. In the helicity basis the $\tau$ pair is hyperentangled in the massless limit, with one ebit in momentum and one in helicity, and the mass-suppressed amplitudes transfer entanglement from within the momentum and spin sectors to between them: at Belle~II, and with our phase convention for the helicity states, the momentum and spin states each have concurrence $0.84$, and the negativity between momenta and spins is $0.19$. The mixedness of the momentum state of Sec.~\ref{sec:momentum} measures this entanglement once the spins are discarded. In the canonical basis, instead, parity gives a momentum Bell state times a separable spin state. Since the spin operators of both bases commute with the momenta, both decompositions are legitimate. The amount of momentum--spin entanglement depends on how the spin is defined at different momenta, and the canonical definition is the one singled out by rotation covariance and parity.

The hemisphere states keep only a part of the produced state: at Belle~II, a fraction $0.86$ in the canonical basis, and $0.31$ or $0.025$ in the helicity basis for hemispheres along $y$ or along the beam. The pseudo-spin operators of Sec.~\ref{sec:pseudospin} keep the whole state and have the measured hemisphere probabilities on the diagonal of the density matrix. In the canonical basis, momentum conservation and parity alone then fix the state of the momenta to be a Bell state, for any choice of hemispheres and at any energy. In the helicity basis the state is a mixture of two Bell states, with concurrence $0.42$ at Belle~II with our phase convention for the helicity states, and $0.53$ with the alternative convention of Sec.~\ref{sec:spin_def}, and for unpolarized beams momenta and spins are separable. The two discretizations answer different questions---which partial waves the pair is in, and how coherent each configuration is with the inverted one---and they agree, because of parity, when the spin is quantized along a fixed axis.

None of the entanglement involving the momenta can be tested at a collider: the measurements are diagonal in the momenta, and all the data are then reproduced by a classical--quantum state, separable between momenta and spins. The spin degrees of freedom, on the other hand, remain fully measurable through the decay products, and spin tomography provides access to genuine spin entanglement. Unobservable as it is, this entanglement alters the picture of what a collider produces: the momenta are not classical attributes that label the events, but quantum variables entangled with each other and with the spins, and they appear classical only because of the limited set of measurements available. For this reason, our analysis has implications for completions of quantum mechanics that rely on deterministic hidden variables: the entanglement of the momenta, together with the Bell nonlocality of the coarse-grained state, is not reproduced by models in which the momenta themselves act as local hidden variables~\cite{kasday1971}, although such models account for all the data available at colliders. Converting this conceptual result into an experimental test necessitates measurements of momenta in a basis of superpositions of hemispheres, which are not accessible at colliders.

\vskip0.3cm
\textit{Acknowledgments---} {\small
LM is supported by the Estonian Research Council under the RVTT3, TK202 and PRG1884 grants.}

\small
\bibliographystyle{JHEP}
\bibliography{bell.bib}

@article{Peres:2002wx,
    author = "Peres, Asher and Scudo, Petra F. and Terno, Daniel R.",
    title = "{Quantum entropy and special relativity}",
    eprint = "quant-ph/0203033",
    archivePrefix = "arXiv",
    journal = "Phys. Rev. Lett.",
    volume = "88",
    pages = "230402",
    year = "2002",
    doi = "10.1103/PhysRevLett.88.230402"
}

@article{Lopes:2015qdr,
    author = "Lopes, Raphael and Imanaliev, Almazbek and Aspect, Alain and Cheneau, Marc and Boiron, Denis and Westbrook, Christoph I.",
    title = "{An atomic Hong-Ou-Mandel experiment}",
    eprint = "1501.03065",
    archivePrefix = "arXiv",
    primaryClass = "quant-ph",
    doi = "10.1038/nature14331",
    month = "1",
    year = "2015"
}

@article{Dussarrat:2017wwi,
    author = "Dussarrat, Pierre and Perrier, Maxime and Imanaliev, Almazbek and Lopes, Raphael and Aspect, Alain and Cheneau, Marc and Boiron, Denis and Westbrook, Christoph",
    title = "{A two-particle, four-mode interferometer for atoms}",
    eprint = "1707.01279",
    archivePrefix = "arXiv",
    primaryClass = "quant-ph",
    doi = "10.1103/PhysRevLett.119.173202",
    journal = "Phys. Rev. Lett.",
    volume = "119",
    pages = "173202",
    year = "2017"
}

@article{Athreya:2025jrg,
    author = "Athreya, Y. S. and Kannan, S. and Yan, X. T. and Lewis-Swan, R. J. and Kheruntsyan, K. V. and Truscott, A. G. and Hodgman, S. S.",
    title = "{Bell correlations between momentum-entangled pairs of $^{4}$He$^{*}$ atoms}",
    eprint = "2502.12392",
    archivePrefix = "arXiv",
    primaryClass = "cond-mat.quant-gas",
    doi = "10.1038/s41467-026-69070-3",
    journal = "Nature Commun.",
    volume = "17",
    number = "1",
    pages = "2357",
    year = "2026"
}

@article{Afik:2025ejh,
    author = "Afik, Yoav and others",
    title = "{Quantum information meets high-energy physics: input to the update of the European strategy for particle physics}",
    eprint = "2504.00086",
    archivePrefix = "arXiv",
    primaryClass = "hep-ph",
    doi = "10.1140/epjp/s13360-025-06752-9",
    journal = "Eur. Phys. J. Plus",
    volume = "140",
    number = "9",
    pages = "855",
    year = "2025"
}

@article{Aguilar-Saavedra:2025njw,
    author = "Aguilar-Saavedra, J. A.",
    title = "{Momentum entanglement at colliders: the $H\to WW,ZZ$ case}",
    eprint = "2512.02104",
    archivePrefix = "arXiv",
    primaryClass = "hep-ph",
    journal = "Phys. Rev. D",
    volume = "114",
    pages = "093002",
    year = "2026"
}

@article{Horodecki:1995nsk,
    author = "Horodecki, R. and Horodecki, P. and Horodecki, M.",
    title = "{Violating Bell inequality by mixed spin-$\frac12$ states: necessary and sufficient condition}",
    journal = "Phys. Lett. A",
    volume = "200",
    pages = "340",
    year = "1995"
}

@article{Rarity:1990zz,
    author = "Rarity, J. G. and Tapster, P. R.",
    title = "{Experimental violation of Bell's inequality based on phase and momentum}",
    journal = "Phys. Rev. Lett.",
    volume = "64",
    pages = "2495",
    year = "1990"
}

@article{Newton:1949cq,
    author = "Newton, T. D. and Wigner, E. P.",
    title = "{Localized states for elementary systems}",
    journal = "Rev. Mod. Phys.",
    volume = "21",
    pages = "400",
    year = "1949"
}

@article{Pryce:1948pf,
    author = "Pryce, M. H. L.",
    title = "{The mass-centre in the restricted theory of relativity and its connexion with the quantum theory of elementary particles}",
    journal = "Proc. Roy. Soc. Lond. A",
    volume = "195",
    pages = "62",
    year = "1948"
}

@article{Fine:1982lwu,
    author = "Fine, Arthur",
    title = "{Hidden Variables, Joint Probability, and the Bell Inequalities}",
    doi = "10.1103/PhysRevLett.48.291",
    journal = "Phys. Rev. Lett.",
    volume = "48",
    number = "5",
    pages = "291",
    year = "1982"
}

@article{Kwiat:1997,
    author = "Kwiat, Paul G.",
    title = "{Hyper-entangled states}",
    journal = "J. Mod. Opt.",
    volume = "44",
    pages = "2173",
    year = "1997"
}

@article{Fabbrichesi:2025aqp,
    author = "Fabbrichesi, M. and Floreanini, R. and Marzola, L.",
    title = "{About Witnessing Bell Non-locality at Colliders}",
    eprint = "2503.18535",
    archivePrefix = "arXiv",
    primaryClass = "quant-ph",
    doi = "10.1007/s10701-025-00894-7",
    journal = "Found. Phys.",
    volume = "55",
    number = "6",
    pages = "83",
    year = "2025"
}

@article{PhysRevLett.60.2731,
  title = {Quantum Correlations of Phase in Nondegenerate Parametric Oscillation},
  author = {Reid, M. D. and Drummond, P. D.},
  journal = {Phys. Rev. Lett.},
  volume = {60},
  issue = {26},
  pages = {2731--2733},
  numpages = {0},
  year = {1988},
  month = {Jun},
  publisher = {American Physical Society},
  doi = {10.1103/PhysRevLett.60.2731},
  url = {https://link.aps.org/doi/10.1103/PhysRevLett.60.2731}
}

@article{PhysRevLett.68.3663,
  title = {Realization of the {E}instein-{P}odolsky-{R}osen paradox for continuous variables},
  author = {Ou, Z. Y. and Pereira, S. F. and Kimble, H. J. and Peng, K. C.},
  journal = {Phys. Rev. Lett.},
  volume = {68},
  issue = {25},
  pages = {3663--3666},
  numpages = {0},
  year = {1992},
  month = {Jun},
  publisher = {American Physical Society},
  doi = {10.1103/PhysRevLett.68.3663},
  url = {https://link.aps.org/doi/10.1103/PhysRevLett.68.3663}
}

@article{PhysRevLett.92.210403,
  title = {Realization of the Einstein-Podolsky-Rosen Paradox Using Momentum- and Position-Entangled Photons from Spontaneous Parametric Down Conversion},
  author = {Howell, John C. and Bennink, Ryan S. and Bentley, Sean J. and Boyd, R. W.},
  journal = {Phys. Rev. Lett.},
  volume = {92},
  issue = {21},
  pages = {210403},
  numpages = {4},
  year = {2004},
  month = {May},
  publisher = {American Physical Society},
  doi = {10.1103/PhysRevLett.92.210403},
  url = {https://link.aps.org/doi/10.1103/PhysRevLett.92.210403}
}

@article{Mancini:2001qbv,
    author = "Mancini, Stefano and Giovannetti, Vittorio and Vitali, David and Tombesi, Paolo",
    title = "{Entangling macroscopic oscillators exploiting radiation pressure}",
    eprint = "quant-ph/0108044",
    archivePrefix = "arXiv",
    doi = "10.1103/PhysRevLett.88.120401",
    journal = "Phys. Rev. Lett.",
    volume = "88",
    pages = "120401",
    year = "2002"
}

@book{Bell_Aspect_2004, 
place={Cambridge}, edition={2}, 
title={Speakable and Unspeakable in Quantum Mechanics: Collected Papers on Quantum Philosophy}, 
publisher={Cambridge University Press}, 
author={Bell, J. S. and Aspect, Alain}, 
year={2004}}

@article{Giese:2025rbx,
    author = "Giese, Enno",
    title = "{Entanglement and Its Verification: A Tutorial on Classical and Quantum Correlations}",
    eprint = "2511.09507",
    archivePrefix = "arXiv",
    primaryClass = "quant-ph",
    month = "11",
    year = "2025"
}

@article{PhysRevA.40.913,
  title = {Demonstration of the Einstein-Podolsky-Rosen paradox using nondegenerate parametric amplification},
  author = {Reid, M. D.},
  journal = {Phys. Rev. A},
  volume = {40},
  issue = {2},
  pages = {913--923},
  numpages = {0},
  year = {1989},
  month = {Jul},
  publisher = {American Physical Society},
  doi = {10.1103/PhysRevA.40.913},
  url = {https://link.aps.org/doi/10.1103/PhysRevA.40.913}
}

@article{Patil:2025jvf,
    author = {Patil, Satyajeet and T{\"o}pfer, Sebastian and Swarnkar, Rajshree and Tovar-Perez, Sergio and Moos, Jonas and Fuenzalida, Jorge and Gr{\"a}fe, Markus},
    title = "{Advances in Position-Momentum Entanglement: A Versatile Tool for Quantum Technologies}",
    eprint = "2505.22265",
    archivePrefix = "arXiv",
    primaryClass = "quant-ph",
    month = "5",
    year = "2025"
}

@article{RevModPhys.75.715,
  title = {Decoherence, einselection, and the quantum origins of the classical},
  author = {Zurek, Wojciech Hubert},
  journal = {Rev. Mod. Phys.},
  volume = {75},
  issue = {3},
  pages = {715--775},
  numpages = {0},
  year = {2003},
  month = {May},
  publisher = {American Physical Society},
  doi = {10.1103/RevModPhys.75.715},
  url = {https://link.aps.org/doi/10.1103/RevModPhys.75.715}
}

@incollection{kasday1971,
      author    = "Kasday, L. R.",
      title     = "{Experimental test of quantum predictions for widely separated photons}",
      booktitle = "{Foundations of Quantum Mechanics, Proceedings of the International School of Physics ``Enrico Fermi'', Course IL}",
      editor    = "d'Espagnat, B.",
      publisher = "Academic Press",
      address   = "New York",
      year      = "1971",
      pages     = "195"
  }

@book{Figari2014,
author = {Figari, R and Teta, A},
	publisher = {Springer Berlin, Heidelberg},
	title = {{Quantum dynamics of a particle in a tracking chamber}},
	year = {2014},
	}

@article{Mott1929,
	author = {Mott, N. F.},
	journal = {Royal Society, Proc.  A},
	pages = {79},
	title = {{The wave mechanics of $\alpha$-ray tracks}},
	volume = {126},
	number={800},
	year = {1929},
	}

@article{Reid:2009zz,
    author = "Reid, M. D. and Drummond, P. D. and Cavalcanti, E. G. and Bowen, W. P. and Lam, P. K. and Bachor, H. A. and Andersen, U. L. and Leuchs, G.",
    title = "{Colloquium: The {E}instein-{P}odolsky-{R}osen paradox: From concepts to applications}",
    eprint = "0806.0270",
    archivePrefix = "arXiv",
    primaryClass = "quant-ph",
    doi = "10.1103/RevModPhys.81.1727",
    journal = "Rev. Mod. Phys.",
    volume = "81",
    pages = "1727--1751",
    year = "2009"
}

@article{Seki:2014cgq,
    author = "Seki, Shigenori and Park, I. Y. and Sin, Sang-Jin",
    title = "{Variation of Entanglement Entropy in Scattering Process}",
    eprint = "1412.7894",
    archivePrefix = "arXiv",
    primaryClass = "hep-th",
    doi = "10.1016/j.physletb.2015.02.028",
    journal = "Phys. Lett. B",
    volume = "743",
    pages = "147--153",
    year = "2015"
}

@article{Peschanski:2016hgk,
    author = "Peschanski, Robi and Seki, Shigenori",
    title = "{Entanglement Entropy of Scattering Particles}",
    eprint = "1602.00720",
    archivePrefix = "arXiv",
    primaryClass = "hep-th",
    doi = "10.1016/j.physletb.2016.04.063",
    journal = "Phys. Lett. B",
    volume = "758",
    pages = "89--92",
    year = "2016"
}

@article{Peschanski:2019yah,
    author = "Peschanski, Robi and Seki, Shigenori",
    title = "{Evaluation of Entanglement Entropy in High Energy Elastic Scattering}",
    eprint = "1906.09696",
    archivePrefix = "arXiv",
    primaryClass = "hep-th",
    doi = "10.1103/PhysRevD.100.076012",
    journal = "Phys. Rev. D",
    volume = "100",
    number = "7",
    pages = "076012",
    year = "2019"
}

@article{Faleiro:2016lsf,
    author = "Faleiro, Ricardo and Costa, Helder A. S. and Pav{\~a}o, Rafael and Hiller, Brigitte and Blin, Alex H. and Sampaio, Marcos",
    title = "{Perturbative approach to entanglement generation in QFT using the S matrix}",
    eprint = "1607.01715",
    archivePrefix = "arXiv",
    primaryClass = "hep-ph",
    doi = "10.1088/1751-8121/aba214",
    journal = "J. Phys. A",
    volume = "53",
    number = "36",
    pages = "365301",
    year = "2020"
}

@article{Kowalska:2024kbs,
    author = "Kowalska, Kamila and Sessolo, Enrico Maria",
    title = "{Entanglement in flavored scalar scattering}",
    eprint = "2404.13743",
    archivePrefix = "arXiv",
    primaryClass = "hep-ph",
    doi = "10.1007/JHEP07(2024)156",
    journal = "JHEP",
    volume = "07",
    pages = "156",
    year = "2024"
}

@article{Low:2024hvn,
    author = "Low, Ian and Yin, Zhewei",
    title = "{Elastic cross section is entanglement entropy}",
    eprint = "2410.22414",
    archivePrefix = "arXiv",
    primaryClass = "hep-th",
    doi = "10.1103/PhysRevD.111.065027",
    journal = "Phys. Rev. D",
    volume = "111",
    number = "6",
    pages = "065027",
    year = "2025"
}

@article{Sou:2025tyf,
    author = "Sou, Chon Man and Wang, Yi and Zhang, Xingkai",
    title = "{Entanglement features in scattering mediated by heavy particles}",
    eprint = "2507.03555",
    archivePrefix = "arXiv",
    primaryClass = "hep-th",
    doi = "10.1007/JHEP10(2025)003",
    journal = "JHEP",
    volume = "10",
    pages = "003",
    year = "2025"
}

@article{Peschanski:2026edo,
    author = "Peschanski, Robi and Seki, Shigenori",
    title = "{Entanglement in Elastic and Inelastic Two-particle Scatterings at High Energy}",
    eprint = "2601.22502",
    archivePrefix = "arXiv",
    primaryClass = "hep-th",
    month = "1",
    year = "2026"
}

@article{Cetto1985,
	author = {Cetto, A. M. and  De la Pe\~{n}a, L. and Santos, E.},
	journal = {Phys. Lett. A},
	pages = {304--306},
	title = {{A Bell inequality involving position, momentum and energy}},
	volume = {113},
	year = {1985},
	}

@article{Belle-II:2018jsg,
	archiveprefix = {arXiv},
	author = {Altmannshofer, W. and others},
	collaboration = {Belle-II},
	doi = {10.1093/ptep/ptz106},
	editor = {Kou, E. and Urquijo, P.},
	eprint = {1808.10567},
	journal = {PTEP},
	note = {[Erratum: PTEP 2020, 029201 (2020)]},
	number = {12},
	pages = {123C01},
	primaryclass = {hep-ex},
	reportnumber = {KEK Preprint 2018-27, BELLE2-PUB-PH-2018-001, FERMILAB-PUB-18-398-T, JLAB-THY-18-2780, INT-PUB-18-047, UWThPh 2018-26},
	title = {{The Belle II Physics Book}},
	volume = {2019},
	year = {2019}}

@article{Horodecki:2009zz,
	archiveprefix = {arXiv},
	author = {Horodecki, Ryszard and Horodecki, Pawel and Horodecki, Michal and Horodecki, Karol},
	doi = {10.1103/RevModPhys.81.865},
	eprint = {quant-ph/0702225},
	journal = {Rev. Mod. Phys.},
	pages = {865--942},
	title = {{Quantum entanglement}},
	volume = {81},
	year = {2009}}

@article{Einstein:1935rr,
	author = {Einstein, Albert and Podolsky, Boris and Rosen, Nathan},
	doi = {10.1103/PhysRev.47.777},
	journal = {Phys. Rev.},
	pages = {777--780},
	title = {{Can quantum mechanical description of physical reality be considered complete?}},
	volume = {47},
	year = {1935}}

@article{Bell:1964,
	author = {Bell, J.S.},
	doi = {10.1103/PhysicsPhysiqueFizika.1.195},
	journal = {Physics Physique Fizika},
	pages = {195},
	publisher = {American Physical Society},
	title = {On the {Einstein} {Podolsky} {Rosen} paradox},
	volume = {1},
	year = {1964}}

@article{Barr:2024djo,
    author = "Barr, Alan J. and Fabbrichesi, Marco and Floreanini, Roberto and Gabrielli, Emidio and Marzola, Luca",
    title = "{Quantum entanglement and Bell inequality violation at colliders}",
    eprint = "2402.07972",
    archivePrefix = "arXiv",
    primaryClass = "hep-ph",
    doi = "10.1016/j.ppnp.2024.104134",
    journal = "Prog. Part. Nucl. Phys.",
    volume = "139",
    pages = "104134",
    year = "2024"
}

@book{Nielsen:2012yss,
	author = {Nielsen, Michael A. and Chuang, Isaac L.},
	doi = {10.1017/cbo9780511976667},
	month = {6},
	publisher = {Cambridge University Press},
	title = {{Quantum Computation and Quantum Information}},
	year = {2012}}

@book{bruss2019quantum,
	author = {Bruss, D. and Leuchs, G.},
	isbn = {9783527805778},
	publisher = {Wiley},
	title = {Quantum Information: From Foundations to Quantum Technology Applications},
	year = {2019}}

@book{Benatti:2010,
	author = {Benatti, F. and Fannes, M. and Floreanini, R. and Petritis, D.},
	doi = {10.1007/978-3-642-11914-9},
	publisher = {Springer Berlin, Heidelberg},
	title = {{Quantum Information, Computation and Cryptography}},
	year = {2010}}

@article{Clauser:1969ny,
	author = {Clauser, John F. and Horne, Michael A. and Shimony, Abner and Holt, Richard A.},
	doi = {10.1103/PhysRevLett.23.880},
	journal = {Phys. Rev. Lett.},
	pages = {880--884},
	title = {{Proposed experiment to test local hidden variable theories}},
	volume = {23},
	year = {1969}}

@article{Fabbrichesi:2024rec,
	archiveprefix = {arXiv},
	author = {Fabbrichesi, M. and Floreanini, R. and Gabrielli, E. and Marzola, L.},
	doi = {10.1103/PhysRevD.110.053008},
	eprint = {2406.17772},
	journal = {Phys. Rev. D},
	number = {5},
	pages = {053008},
	primaryclass = {hep-ph},
	title = {{Bell inequality is violated in charmonium decays}},
	volume = {110},
	year = {2024}}

@article{Fabbrichesi:2023idl,
	archiveprefix = {arXiv},
	author = {Fabbrichesi, M. and Floreanini, R. and Gabrielli, E. and Marzola, L.},
	doi = {10.1103/PhysRevD.109.L031104},
	eprint = {2305.04982},
	journal = {Phys. Rev. D},
	number = {3},
	pages = {L031104},
	primaryclass = {hep-ph},
	title = {{Bell inequality is violated in $B^{0}\to J/\psi\,K^{*0}(892)$ decays}},
	volume = {109},
	year = {2024}}

@article{CMS:2024pts,
	archiveprefix = {arXiv},
	author = {Hayrapetyan, Aram and others},
	collaboration = {CMS},
	doi = {10.1088/1361-6633/ad7e4d},
	eprint = {2406.03976},
	journal = {Rept. Prog. Phys.},
	number = {11},
	pages = {117801},
	primaryclass = {hep-ex},
	reportnumber = {CMS-TOP-23-001, CERN-EP-2024-137},
	title = {{Observation of quantum entanglement in top quark pair production in proton\textendash{}proton collisions at $\sqrt{s} = 13$ TeV}},
	volume = {87},
	year = {2024}}

@article{ATLAS:2023fsd,
	archiveprefix = {arXiv},
	author = {Aad, Georges and others},
	collaboration = {ATLAS},
	doi = {10.1038/s41586-024-07824-z},
	eprint = {2311.07288},
	journal = {Nature},
	number = {8030},
	pages = {542--547},
	primaryclass = {hep-ex},
	reportnumber = {CERN-EP-2023-230},
	title = {{Observation of quantum entanglement with top quarks at the ATLAS detector}},
	volume = {633},
	year = {2024}}

@article{Belle-II:2024vuc,
    author = "Adachi, I. and others",
    collaboration = "Belle-II",
    title = "{Measurement of the integrated luminosity of data samples collected during 2019-2022 by the Belle II experiment*}",
    eprint = "2407.00965",
    archivePrefix = "arXiv",
    primaryClass = "hep-ex",
    reportNumber = "Belle II Preprint 2024-019; KEK Preprint 2024-16",
    doi = "10.1088/1674-1137/ad806c",
    journal = "Chin. Phys. C",
    volume = "49",
    number = "1",
    pages = "013001",
    year = "2025"
}

@article{Chen:2002elr,
    author = "Chen, Zeng-Bing and Pan, Jian-Wei and Hou, Guang and Zhang, Yong-De",
    title = "{Maximal Violation of Bell's Inequalities for Continuous Variable Systems}",
    eprint = "quant-ph/0103051",
    archivePrefix = "arXiv",
    doi = "10.1103/PhysRevLett.88.040406",
    journal = "Phys. Rev. Lett.",
    volume = "88",
    pages = "040406",
    year = "2002"
}

@article{Gour:2003wsm,
    author = "Gour, G. and Khanna, F. C. and Mann, A. and Revzen, M.",
    title = "{Optimization of Bell's Inequality Violation For Continuous Variable Systems}",
    eprint = "quant-ph/0308063",
    archivePrefix = "arXiv",
    doi = "10.1016/j.physleta.2004.03.018",
    month = "10",
    year = "2003"
}

@article{Banaszek:1998wcc,
    author = "Banaszek, Konrad and W{\'o}dkiewicz, Krzysztof",
    title = "{Testing Quantum Nonlocality in Phase Space}",
    eprint = "quant-ph/9806068",
    archivePrefix = "arXiv",
    doi = "10.1103/PhysRevLett.82.2009",
    journal = "Phys. Rev. Lett.",
    volume = "82",
    number = "10",
    pages = "2009",
    year = "1999"
}

\end{document}